\documentclass[11pt,a4paper]{article}

\usepackage[utf8]{inputenc}
\usepackage[T1]{fontenc}
\usepackage{amsmath,amsthm,amssymb}
\usepackage{algorithm}
\usepackage{algpseudocode}
\usepackage{booktabs}
\usepackage{array}
\usepackage{xcolor}
\usepackage{natbib}
\usepackage{hyperref}
\usepackage{geometry}
\usepackage{enumitem}
\usepackage{pifont}  
\usepackage{tikz}
\usetikzlibrary{shapes.geometric,positioning}
\usepackage{listings}

\definecolor{codegreen}{rgb}{0.13,0.55,0.13}      
\definecolor{codegray}{rgb}{0.5,0.5,0.5}
\definecolor{codepurple}{rgb}{0.58,0,0.82}
\definecolor{codeblue}{rgb}{0.0,0.0,0.8}          
\definecolor{backcolour}{rgb}{0.95,0.95,0.95}
\definecolor{codestring}{rgb}{0.64,0.08,0.08}     

\lstdefinestyle{pythonstyle}{
    language=Python,
    commentstyle=\color{codegreen},
    keywordstyle=\color{codeblue},
    stringstyle=\color{codestring},
}

\lstdefinelanguage{Dafny}{
    keywords={method, function, predicate, lemma, class, datatype, type, var, const,
              if, else, while, for, forall, exists, in, return, returns, requires,
              ensures, invariant, decreases, modifies, reads, assert, assume, calc,
              match, case, new, this, null, true, false, old, fresh, ghost, static,
              import, module, export, abstract, trait, extends, implements,
              int, nat, bool, real, string, char, array, seq, set, multiset, map,
              object, ORDINAL},
    keywordstyle=\color{codeblue},
    sensitive=true,
    comment=[l]{//},
    morecomment=[s]{/*}{*/},
    commentstyle=\color{codegreen},
    stringstyle=\color{codestring},
    morestring=[b]",
    morestring=[b]'
}

\lstdefinestyle{dafnystyle}{
    language=Dafny,
    basewidth=0.5em
}

\usepackage{tcolorbox}

\newtheorem{invariant}{Loop Invariant}
\newtheorem{postcondition}{Post-condition}
\newcommand{\urs}{\;\texttt{>}\texttt{>}\texttt{>}\;}

\algrenewcommand{\algorithmicindent}{2\fontcharwd\font`\ }  

\definecolor{inclusive}{RGB}{0,100,0}
\definecolor{exclusive}{RGB}{180,0,0}
\definecolor{highlight}{RGB}{255,255,200}

\tcbset{
    invbox/.style={
        colback=blue!5!white,
        colframe=blue!75!black,
        fonttitle=\bfseries,
        title=#1
    }
}

\title{\textbf{Binary Search Variants: A Comprehensive Analysis}\\[0.5em]
  \large Invariants, Boundary Conventions, and Implementation Pitfalls}

\author{Ali Dasdan\thanks{The author acknowledges the use of AI
    assistants (Claude, Gemini) for language editing, reference
    checking, and debugging. The scientific content,
    analysis, and conclusions are exclusively the author's own, as is
    responsibility for any errors.}\\
  KD Consulting\\ Saratoga, CA,
  USA\\
  alidasdan@gmail.com\\}

\date{}

\begin{document}

\maketitle

\begin{abstract}
Binary search is deceptively simple in concept yet notoriously
difficult to implement correctly. This paper presents a unified
treatment of binary search: five core variants, six derived query
functions, and four standard library implementations (BSD, glibc, Java,
C++ STL), each with consistent notation, loop invariants, and
analysis. We introduce \texttt{bsearch\_ultimate}, a combined search
that subsumes all variants in a single call. Every algorithm is
provided as synchronized Python code, Dafny formal proof, and
pseudocode. All implementations are validated by over 9,500 tests and
21 Dafny formal verifications; an additional six deliberately faulty
implementations demonstrate common bug categories and Dafny's ability
to detect them.  We also provide memorable rules linking boundary
choices to loop conditions and update formulas.
\end{abstract}

\tableofcontents

\section{Introduction}

Suppose we are given a sorted array $A$ of $n$ elements and we are
asked to determine whether $A$ contains a given element $t$ (the
\emph{target}). A linear scan solves this in $O(n)$ time by comparing
each element to $t$. \emph{Binary search} does better: because $A$ is
sorted, comparing $t$ to the middle element tells us which half of the
array can contain $t$, letting us discard the other half. Repeating
this halving yields an $O(\lg n)$ algorithm.

\subsection{Historical Notes}

Binary search is deceptively simple in concept yet notoriously
difficult to implement correctly. The first mention of binary search
in computing was by John Mauchly in 1946 as part of the Moore School
Lectures. Remarkably, every published binary search algorithm worked
only for arrays whose length was one less than a power of two
($2^n - 1$) until Derrick Lehmer published a general version in 1960
\citep{knuth1998}. \citet{bottenbruch1962} introduced the
deferred-equality variant in 1962. \citet{knuth1998} also describes a
\emph{uniform binary search} variant using precomputed deltas instead
of bounds, designed for machines where index arithmetic was expensive;
this optimization is obsolete on modern CPUs. Binary search also
survives as a subroutine inside modern algorithms; for example, the
galloping merge step of Timsort, Python's standard sorting algorithm,
relies on it \citep{timsort}.

The algorithm's difficulty is well documented. In courses taught at IBM
and Bell Labs, \citet{bentley1986} asked professional programmers to
implement binary search as an in-class exercise. When given thirty
minutes to test and fix their own code, roughly 90\% still had bugs;
Bentley expressed doubt that the remaining 10\% were correct either. A
1988 study found correct binary search code in only 5 of 20 textbooks
\citep{knuth1998}. The subtlety extends even to experts: Bentley's own
implementation in \textit{Programming Pearls} contained an integer
overflow bug that remained undetected for over twenty years, and Java's
standard library carried the same bug for nine years
\citep{wikipedia_bsearch}.

\subsection{Classification by Comparison Strategy}

Although the basic idea is simple---compare the target to the middle
element and recurse on one half---there is no single binary search
algorithm. \emph{Binary search variants} differ along several axes
(detailed in the next subsection) but can be broadly classified into
two fundamental categories based on how they handle comparisons:

\begin{enumerate}
    \item \textbf{3-Way Comparison}
      (Sections~\ref{sec:variant1}--\ref{sec:variant2}): These algorithms
      check for equality \emph{inside} the loop using three branches
      on the midpoint element $A[m]$: $A[m] = t$, $A[m] < t$, and
      $A[m] > t$. They return immediately upon finding the
      target. Best for finding \emph{any} matching element.

    \item \textbf{2-Way Comparison}
      (Sections~\ref{sec:variant3}--\ref{sec:variant4}): These algorithms
      use only two branches inside the loop---$A[m] < t$ or
      $A[m] \geq t$---folding equality into one side and deferring the
      exact match check until \emph{after} the loop terminates. They
      are essential for finding the \emph{leftmost} or
      \emph{rightmost} occurrence and are robust when duplicates are
      present.
\end{enumerate}

\noindent A third category involves \textbf{alternative midpoint
  calculations} (Section~\ref{sec:variant5}) to avoid infinite loops in
certain update patterns.

\subsection{Implementation Details}

The binary search variants in the previous subsection arise from
different choices along the following five axes, and the difficulty of
getting these details right is what makes binary search so
error-prone:

\begin{enumerate}[noitemsep]
    \item \textbf{Boundary conventions}: Are the left and right bounds inclusive or exclusive?
    \item \textbf{Loop termination}: When does the search conclude?
    \item \textbf{Mid-point calculation}: How is the midpoint computed to avoid overflow?
    \item \textbf{Bound updates}: How are $l$ and $r$ updated after each comparison?
    \item \textbf{Return semantics}: What is returned when the target is found or not found?
\end{enumerate}

\begin{tcolorbox}[invbox={The Golden Rule for Loop Conditions}]
\textbf{If the right bound is exclusive, use $l < r$}.\\
\textbf{If the right bound is inclusive, use $l \leq r$}.\\[0.5em]
This rule ensures the search space is non-empty when entering the loop.
\end{tcolorbox}

\subsection{Contributions}

This paper makes the following contributions:

\begin{enumerate}[noitemsep]
    \item \textbf{Unified treatment of five core variants}: Standard
      search with $[l,r)$ and $[l,r]$ bounds, leftmost match,
      rightmost match, and Bottenbruch's deferred-equality algorithm,
      each with consistent notation, loop invariants, and analysis.

    \item \textbf{Six derived query functions}: We show how
      \texttt{bisect\_left} and \texttt{bisect\_right} yield
      \texttt{find\_rank}, \texttt{find\_pred\_strict},
      \texttt{find\_succ\_strict}, \texttt{find\_floor},
      \texttt{find\_ceil}, and \texttt{find\_range} as thin wrappers.

    \item \textbf{Standard library survey}: We analyze the BSD
      (Apple), glibc, Java (OpenJDK), and C++ STL implementations,
      mapping each to our core variants.

    \item \textbf{Combined search (\texttt{bsearch\_ultimate})}: A
      single function returning leftmost index, rightmost index, and
      insertion point, subsuming all other variants and derived
      functions (Section~\ref{sec:universality}).

    \item \textbf{Synchronized triple implementation}: Every algorithm
      is provided as Python code, a Dafny formal proof, and
      pseudocode, kept in strict correspondence.

    \item \textbf{Comprehensive validation}: Over 9,500 Python tests
      and 21 Dafny formal verifications confirm correctness; 6
      deliberately faulty implementations demonstrate common bug
      categories.

    \item \textbf{Practical guidance}: A decision guide, common
      pitfalls, and memorable rules linking boundary conventions to
      loop conditions and update formulas.
\end{enumerate}

\subsection{Paper Organization}

Section~\ref{sec:notation} establishes notation and conventions used
throughout.
Sections~\ref{sec:variant1}--\ref{sec:variant5} present the five core
binary search variants, each with a loop invariant proof, pseudocode,
Python implementation, and formally verified Dafny implementation.
Section~\ref{sec:derived} derives six approximate-match query functions
(rank, predecessor, successor, floor, ceiling, range count) from
\texttt{bisect\_left} and \texttt{bisect\_right}.
Section~\ref{sec:bisect} connects these to Python's \texttt{bisect} module,
and Section~\ref{sec:stdlib} surveys four standard library implementations
(BSD, glibc, Java, C++ STL), mapping each to our core variants.

Section~\ref{sec:ultimate} introduces \texttt{bsearch\_ultimate}, a combined
search returning leftmost index, rightmost index, and insertion point
in a single call, presented in three formulations (self-contained,
compositional, and count-based) with a universality proof
(Section~\ref{sec:universality}) showing it subsumes all other variants.
Section~\ref{sec:comparison} provides a comprehensive comparison table,
Section~\ref{sec:pitfalls} catalogs common pitfalls, and
Section~\ref{sec:decision} offers a decision guide for variant selection.

Section~\ref{sec:testing} describes our testing strategy (9,566 tests
across ten suites), Section~\ref{sec:performance} presents performance
benchmarks in both Python and C revealing language-dependent ranking
reversals, and Section~\ref{sec:dafny} details the Dafny formal
verification of all 21 correct implementations and 6 deliberately
faulty ones.  Section~\ref{sec:conclusion} concludes.

\section{Notation and Conventions}
\label{sec:notation}

Throughout this paper, we use the following notation:
\begin{itemize}[noitemsep]
    \item $A[0..n-1]$: A sorted array of $n$ elements in non-decreasing order
    \item $t$: The target value being searched
    \item $l, r$: Left and right bounds of the current search interval
    \item $m$: The midpoint index
    \item $[\textcolor{inclusive}{\text{inclusive}}]$: The bound includes the endpoint
    \item $(\textcolor{exclusive}{\text{exclusive}})$: The bound excludes the endpoint
\end{itemize}

\noindent\textbf{Search interval notation}:
\begin{itemize}[noitemsep]
    \item $[l, r)$: $l$ inclusive, $r$ exclusive --- valid indices are $\{l, l+1, \ldots, r-1\}$
    \item $[l, r]$: Both inclusive --- valid indices are $\{l, l+1, \ldots, r\}$
\end{itemize}

\section{Variant 1: Standard Binary Search with \texorpdfstring{$[l, r)$}{[l, r)} Convention}
\label{sec:variant1}

\noindent\textit{Category: 3-Way Comparison (equality check inside loop)}

\medskip
This variant uses left-inclusive, right-exclusive bounds---the
convention used by Python's \texttt{range()}, C++'s iterators, and
many algorithms textbooks \citep{knuth1998, cormen2009, cmu15122}.

\subsection{Algorithm}

\begin{algorithm}[H]
\caption{Binary Search with $[l, r)$ bounds}
\begin{algorithmic}[1]
\Function{bsearch1}{$A, t$}
    \State $l \gets 0$
    \State $r \gets n$ \Comment{$r$ is exclusive: valid range is $[0, n)$}
    \While{$l < r$} \Comment{Search space non-empty iff $l < r$}
        \State $m \gets l + \lfloor(r - l) / 2\rfloor$
        \If{$A[m] = t$}  \Comment{Target found}
            \State \Return $m$
        \ElsIf{$A[m] < t$}
            \State $l \gets m + 1$ \Comment{Target in $[m+1, r)$}
        \Else
            \State $r \gets m$ \Comment{Target in $[l, m)$; exclude $m$}
        \EndIf
    \EndWhile
    \State \Return $-1$  \Comment{Target not found}
\EndFunction
\end{algorithmic}
\end{algorithm}

\subsection{Loop Invariant}
\label{sec:loop-invariant1}

\begin{invariant}[bsearch1]
At the start of each iteration:
\begin{enumerate}[noitemsep]
    \item The search interval $[l, r)$ contains all possible locations where $t$ could be found
    \item $0 \leq l \leq r \leq n$
    \item For all $i < l$: $A[i] < t$ (everything left of $l$ is too small)
    \item For all $i \geq r$: $A[i] > t$ (everything at or right of $r$ is too large)
\end{enumerate}
\end{invariant}

\subsection{Key Properties}

\begin{itemize}[noitemsep]
    \item \textbf{Initialization}: $l = 0$, $r = n$ gives $[0, n)$, the entire array
    \item \textbf{Termination}: Loop exits when $l \geq r$, meaning $[l, r) = \emptyset$
    \item \textbf{Why $r \gets m$}: Since $r$ is exclusive and $A[m] \neq t$ (and $A[m] > t$), we exclude $m$ by setting $r = m$
    \item \textbf{Why $l \gets m + 1$}: Since $l$ is inclusive and $A[m] < t$, we must skip past $m$
\end{itemize}

\subsection{Python Implementation}

\begin{lstlisting}[style=pythonstyle, caption=Variant 1: Half-open interval {$[l, r)$}]
def bsearch1(A: list[int], t: int) -> int:
  l, r = 0, len(A)          # r is exclusive
  while l < r:              # [l, r) non-empty iff l < r
    m = l + (r - l) // 2    # floor: l + floor((r-l)/2)
    if A[m] == t:           # Target found
      return m
    elif A[m] < t:
      l = m + 1             # Target in [m+1, r)
    else:
      r = m                 # Target in [l, m)
  return -1                 # Target not found
\end{lstlisting}

\subsection{Dafny Implementation (Formally Verified)}

Each variant in this paper is accompanied by an implementation in
Dafny~\citep{dafny}, a verification-aware programming language
developed at Microsoft Research. Dafny's verifier automatically proves
that an implementation satisfies its specification for \emph{all}
possible inputs---not just the inputs covered by tests. The
specification consists of \emph{preconditions}
(\texttt{requires}~clauses, stating what must hold when the method is
called), \emph{postconditions} (\texttt{ensures}~clauses, stating what
the method guarantees upon return), \emph{loop invariants}
(\texttt{invariant}~clauses, stating properties maintained at each
iteration), and a \texttt{decreases}~clause proving termination.

In the listing below:
\begin{itemize}[noitemsep, topsep=0pt]
  \item \textbf{Preconditions}: These require that $A$ is sorted and
    its length fits within a 32-bit signed integer.
  \item \textbf{Postconditions}: These ensure that if the returned
    index is not $-1$, then $A[\mathit{idx}] = t$; otherwise, $t$ does
    not appear anywhere in the array.
  \item \textbf{Loop Invariants}: These formalize the narrowing
    argument from Section~\ref{sec:loop-invariant1}: all elements to
    the left of~$l$ are less than~$t$, and all elements at or to the
    right of~$r$ are greater than~$t$.
  \item \textbf{Arithmetic Safety}: Finally, the \texttt{assert}
    statement inside the loop is critical. It compels the verifier to
    prove that the midpoint calculation $l + (r - l) / 2$ never
    exceeds the maximum 32-bit signed integer. By successfully
    verifying this assertion, Dafny provides a mathematical guarantee
    that this implementation is immune to the famous integer overflow
    bug, leveraging the precondition that the array length itself fits
    within signed 32-bit bounds.
\end{itemize}

Dafny's automated theorem prover (Z3 \citep{z3}) checks these
conditions exhaustively at compile time. Section~\ref{sec:dafny}
provides a full discussion of the verification methodology and results
for all implementations.

\begin{lstlisting}[style=dafnystyle, caption=Variant 1 in Dafny]
method bsearch1(A: array<int>, t: int) returns (idx: int)
  requires Sorted(A)
  requires A.Length <= MAX_INT32
  ensures -1 <= idx < A.Length
  ensures idx != -1 ==> A[idx] == t
  ensures idx == -1 ==> forall k :: 0 <= k < A.Length ==> A[k] != t
{
  var l := 0;
  var r := A.Length;           // r is exclusive
  while l < r                  // [l, r) non-empty iff l < r
    invariant 0 <= l <= r <= A.Length
    invariant forall k :: 0 <= k < l ==> A[k] < t
    invariant forall k :: r <= k < A.Length ==> A[k] > t
    decreases r - l
  {
    assert l + (r - l) / 2 <= MAX_INT32;  // Overflow check
    var m := l + (r - l) / 2;             // Overflow-safe midpoint
    if A[m] == t {             // Target found
      return m;
    } else if A[m] < t {       // Target in [m+1, r)
      l := m + 1;  
    } else {                   // Target in [l, m)
      r := m;
    }                   
  }
  return -1;                   // Target not found
}
\end{lstlisting}

\section{Variant 2: Bentley's Binary Search with \texorpdfstring{$[l, r]$}{[l, r]} Convention}
\label{sec:variant2}

\noindent\textit{Category: 3-Way Comparison (equality check inside loop)}

\medskip
The classic version of \citet{bentley1986} uses both bounds
inclusive, which is perhaps more intuitive but requires careful
handling.

\subsection{Algorithm}

\begin{algorithm}[H]
\caption{Bentley's Binary Search with $[l, r]$ bounds}
\begin{algorithmic}[1]
\Function{bsearch2}{$A, t$}
    \State $l \gets 0$
    \State $r \gets n - 1$ \Comment{$r$ is inclusive: valid range is $[0, n-1]$}
    \While{$l \leq r$} \Comment{Search space non-empty iff $l \leq r$}
        \State $m \gets l + \lfloor(r - l) / 2\rfloor$
        \If{$A[m] = t$}  \Comment{Target found}
            \State \Return $m$
        \ElsIf{$A[m] < t$}
            \State $l \gets m + 1$  \Comment{Target in $[m+1,r]$}
        \Else
            \State $r \gets m - 1$  \Comment{Target in $[l,m-1]$}
        \EndIf
    \EndWhile
    \State \Return $-1$  \Comment{Target not found}
\EndFunction
\end{algorithmic}
\end{algorithm}

\subsection{Loop Invariant}

\begin{invariant}[bsearch2]
At the start of each iteration:
\begin{enumerate}[noitemsep]
    \item The search interval $[l, r]$ contains all possible locations where $t$ could be found
    \item $0 \leq l$ and $r \leq n - 1$ (with possible $l > r$ indicating empty interval)
    \item For all $i < l$: $A[i] < t$
    \item For all $i > r$: $A[i] > t$
\end{enumerate}
\end{invariant}

\subsection{Critical Differences from Variant 1}

\begin{center}
\begin{tabular}{lcc}
\toprule
\textbf{Aspect} & \textbf{bsearch1 $[l,r)$} & \textbf{bsearch2 $[l,r]$} \\
\midrule
Initial $r$ & $n$ & $n - 1$ \\
Loop condition & $l < r$ & $l \leq r$ \\
Update when $A[m] > t$ & $r \gets m$ & $r \gets m - 1$ \\
Empty interval & $l \geq r$ & $l > r$ \\
\bottomrule
\end{tabular}
\end{center}

\subsection{Python Implementation}

\begin{lstlisting}[style=pythonstyle, caption=Variant 2: Closed interval {$[l, r]$} (Bentley)]
def bsearch2(A: list[int], t: int) -> int:
  l, r = 0, len(A) - 1    # Both bounds inclusive; r = -1 if empty
  while l <= r:           # [l, r] non-empty iff l <= r
    m = l + (r - l) // 2  # floor: l + floor((r-l)/2)
    if A[m] == t:         # Target found
      return m
    elif A[m] < t:        # Target in [m+1, r]
      l = m + 1
    else:
      r = m - 1           # Target in [l, m-1]
  return -1               # Target not found
\end{lstlisting}

\subsection{Dafny Implementation (Formally Verified)}

The Dafny version matches the Python implementation, returning $-1$ if not found:

\begin{lstlisting}[style=dafnystyle, caption=Variant 2 in Dafny]
method bsearch2(A: array<int>, t: int) returns (idx: int)
  requires Sorted(A)
  requires A.Length <= MAX_INT32
  ensures -1 <= idx < A.Length
  ensures idx != -1 ==> A[idx] == t
  ensures idx == -1 ==> forall k :: 0 <= k < A.Length ==> A[k] != t
{
  var l := 0;
  var r := A.Length - 1;  // Both bounds inclusive
  while l <= r
    invariant 0 <= l <= A.Length
    invariant -1 <= r < A.Length
    invariant l <= r + 1
    invariant forall k :: 0 <= k < l ==> A[k] < t
    invariant forall k :: r < k < A.Length ==> A[k] > t
    decreases r - l
  {
    assert l + (r - l) / 2 <= MAX_INT32;  // Overflow check
    var m := l + (r - l) / 2;             // Overflow-safe midpoint
    if A[m] == t {        // Target found
      return m;
    } else if A[m] < t {  // Target in [m+1,r]
      l := m + 1;
    } else {              // Target in [l, m-1]
      r := m - 1;
    }  
  }
  return -1;              // Target not found
}
\end{lstlisting}

\section{Variant 3: Find Leftmost Match (Lower Bound)}
\label{sec:variant3}

\noindent\textit{Category: 2-Way Comparison (equality check deferred; robust with duplicates)}

\medskip
When duplicates exist, we may want the \emph{first} occurrence of
$t$. This variant finds the leftmost index $i$ such that $A[i] = t$,
or $-1$ if $t$ is not present.

\subsection{Algorithm}

\begin{algorithm}[H]
\caption{Find Leftmost Match with $[l, r)$ bounds}
\begin{algorithmic}[1]
\Function{bsearch3}{$A, t$}
    \State $l \gets 0$
    \State $r \gets n$  \Comment{$r$ is exclusive: valid range is $[0, n)$}
    \While{$l < r$}  \Comment{Search space non-empty iff $l < r$}
        \State $m \gets l + \lfloor(r - l) / 2\rfloor$
        \If{$A[m] < t$}  \Comment{Target in $[m+1,r)$}
            \State $l \gets m + 1$
        \Else
            \State $r \gets m$ \Comment{Even if $A[m] = t$, keep searching left}
        \EndIf
    \EndWhile
    \If{$l < n$ \textbf{and} $A[l] = t$} \Comment{Bounds check required!}
        \State \Return $l$  \Comment{Target found}
    \EndIf
    \State \Return $-1$  \Comment{Target not found}
\EndFunction
\end{algorithmic}
\end{algorithm}

\subsection{Loop Invariant}

\begin{invariant}[bsearch3]
At the start of each iteration:
\begin{enumerate}[noitemsep]
    \item All elements $A[0..l-1]$ are strictly less than $t$
    \item All elements $A[r..n-1]$ are greater than or equal to $t$
    \item If $t$ exists in $A$, its leftmost occurrence is in $[l, r)$
\end{enumerate}
\end{invariant}

\subsection{Post-condition}

\begin{postcondition}
When the loop terminates ($l = r$):
\begin{itemize}[noitemsep]
    \item $l$ is the smallest index such that $A[l] \geq t$ (if such index exists)
    \item This is equivalent to Python's \texttt{bisect\_left} behavior
    \item We must explicitly check if $A[l] = t$ since the loop doesn't check equality
\end{itemize}
\end{postcondition}

\subsection{Why No Equality Check in Loop?}

The key insight is that when we find $A[m] = t$, we \emph{don't return
immediately}. Instead, we set $r = m$ to continue searching for an
earlier occurrence. This ensures we find the \emph{leftmost} match.

\begin{tcolorbox}[colback=yellow!10, colframe=orange!75!black]
\textbf{Bug Alert}: The post-loop check \texttt{if A[l] == t} can
cause an index-out-of-bounds error if $l = n$. Always check $l < n$
first!
\end{tcolorbox}

\subsection{Python Implementation}

\begin{lstlisting}[style=pythonstyle, caption=Variant 3: Leftmost match / lower bound]
def bsearch3(A: list[int], t: int) -> int:
  l, r = 0, len(A)          # r is exclusive
  while l < r:
    m = l + (r - l) // 2    # floor: l + floor((r-l)/2)
    if A[m] < t:
      l = m + 1             # Target in [m+1,r)
    else:
      r = m                 # Even if A[m] == t, keep searching left
  # Post-loop: l is insertion point (smallest i where A[i] >= t)
  if l < len(A) and A[l] == t:  # Bounds check required!
    return l                # Target found
  return -1                 # Target not found
\end{lstlisting}

\subsection{Dafny Implementation (Formally Verified)}

The Dafny version matches the Python implementation, returning $-1$ if
not found:

\begin{lstlisting}[style=dafnystyle, caption=Variant 3 in Dafny (leftmost match)]
method bsearch3(A: array<int>, t: int) returns (idx: int)
  requires Sorted(A)
  requires A.Length <= MAX_INT32
  ensures -1 <= idx < A.Length
  ensures idx != -1 ==> A[idx] == t
  ensures idx != -1 ==> (idx == 0 || A[idx-1] < t)  // Leftmost
  ensures idx == -1 ==> forall k :: 0 <= k < A.Length ==> A[k] != t
{
  var l := 0;
  var r := A.Length;      // r is exclusive
  while l < r
    invariant 0 <= l <= r <= A.Length
    invariant forall k :: 0 <= k < l ==> A[k] < t
    invariant forall k :: r <= k < A.Length ==> A[k] >= t
    decreases r - l
  {
    assert l + (r - l) / 2 <= MAX_INT32;  // Overflow check
    var m := l + (r - l) / 2;             // Overflow-safe midpoint
    if A[m] < t {  
      l := m + 1;         // Target in [m+1,r)
    } else {              // Even if A[m] == t, keep searching left
      r := m;             // Target in [l,m)
    }  
  }
  // Post-loop: l is insertion point
  if l < A.Length && A[l] == t {          // Bounds check required!
    return l;             // Target found
  }
  return -1;              // Target not found
}
\end{lstlisting}

\section{Variant 4: Find Rightmost Match (Upper Bound)}
\label{sec:variant4}

\noindent\textit{Category: 2-Way Comparison (equality check deferred; robust with duplicates)}

\medskip
This variant finds the \emph{last} occurrence of $t$.

\subsection{Algorithm}

\begin{algorithm}[H]
\caption{Find Rightmost Match with $[l, r)$ bounds}
\begin{algorithmic}[1]
\Function{bsearch4}{$A, t$}
    \State $l \gets 0$
    \State $r \gets n$  \Comment{$r$ is exclusive: valid range is $[0, n)$}
    \While{$l < r$}
        \State $m \gets l + \lfloor(r - l) / 2\rfloor$
        \If{$A[m] > t$}
            \State $r \gets m$  \Comment{Target in $[l,m)$}
        \Else
            \State $l \gets m + 1$ \Comment{Even if $A[m] = t$, keep searching right}
        \EndIf
    \EndWhile
    \If{$r > 0$ \textbf{and} $A[r-1] = t$} \Comment{Check element just before $r$}
        \State \Return $r - 1$  \Comment{Target found}
    \EndIf
    \State \Return $-1$  \Comment{Target not found}
\EndFunction
\end{algorithmic}
\end{algorithm}

\subsection{Loop Invariant}

\begin{invariant}[bsearch4]
At the start of each iteration:
\begin{enumerate}[noitemsep]
    \item All elements $A[0..l-1]$ are less than or equal to $t$
    \item All elements $A[r..n-1]$ are strictly greater than $t$
    \item If $t$ exists in $A$, its rightmost occurrence is in $[l, r)$
\end{enumerate}
\end{invariant}

\subsection{Post-condition}

\begin{postcondition}
When the loop terminates:
\begin{itemize}[noitemsep]
    \item $l = r$ is the smallest index such that $A[l] > t$
    \item This is equivalent to Python's \texttt{bisect\_right} behavior
    \item The rightmost match (if any) is at index $r - 1$
\end{itemize}
\end{postcondition}

\subsection{Symmetry with Variant 3}

\begin{center}
\begin{tabular}{lcc}
\toprule
\textbf{Aspect} & \textbf{Leftmost (bsearch3)} & \textbf{Rightmost (bsearch4)} \\
\midrule
Comparison in ``else'' & $A[m] \geq t \Rightarrow r = m$ & $A[m] \leq t \Rightarrow l = m + 1$ \\
Final check index & $l$ & $r - 1$ \\
Equivalent to & \texttt{bisect\_left} & \texttt{bisect\_right} $- 1$ \\
\bottomrule
\end{tabular}
\end{center}

\subsection{Python Implementation}

\begin{lstlisting}[style=pythonstyle, caption=Variant 4: Rightmost match / upper bound]
def bsearch4(A: list[int], t: int) -> int:
  l, r = 0, len(A)        # r is exclusive
  while l < r:
    m = l + (r - l) // 2  # floor: l + floor((r-l)/2)
    if A[m] > t:
      r = m               # Target in [l,m)
    else:
      l = m + 1           # Even if A[m] == t, keep searching right
  # Post-loop: r is insertion point (smallest i where A[i] > t)
  if r > 0 and A[r - 1] == t:  # Check element just before r
    return r - 1          # Target found
  return -1               # Target not found
\end{lstlisting}

\subsection{Dafny Implementation (Formally Verified)}

The Dafny version matches the Python implementation, returning $-1$ if not found:

\begin{lstlisting}[style=dafnystyle, caption=Variant 4 in Dafny (rightmost match)]
method bsearch4(A: array<int>, t: int) returns (idx: int)
  requires Sorted(A)
  requires A.Length <= MAX_INT32
  ensures -1 <= idx < A.Length
  ensures idx != -1 ==> A[idx] == t
  ensures idx != -1 ==> (idx == A.Length - 1 || A[idx+1] > t)  // Rightmost
  ensures idx == -1 ==> forall k :: 0 <= k < A.Length ==> A[k] != t
{
  var l := 0;
  var r := A.Length;      // r is exclusive
  while l < r
    invariant 0 <= l <= r <= A.Length
    invariant forall k :: 0 <= k < l ==> A[k] <= t
    invariant forall k :: r <= k < A.Length ==> A[k] > t
    decreases r - l
  {
    assert l + (r - l) / 2 <= MAX_INT32;  // Overflow check
    var m := l + (r - l) / 2;             // Overflow-safe midpoint
    if A[m] > t {
      r := m;             // Target in [l,m)
    } else {              // Even if A[m] == t, keep searching right
      l := m + 1;         // Target in [m+1,r)
    }  
  }
  // Post-loop: r is insertion point
  if r > 0 && A[r - 1] == t {
    return r - 1;         // Target found
  }
  return -1;              // Target not found
}
\end{lstlisting}

\section{Variant 5: Bottenbruch's Algorithm (Deferred Equality)}
\label{sec:variant5}

\noindent\textit{Category: 2-Way Comparison with alternative midpoint calculation}

\medskip
\citeauthor{bottenbruch1962}'s 1962 algorithm \citep{bottenbruch1962,
  wikipedia_bsearch} defers the equality check until after the loop,
using both bounds inclusive.

\subsection{Algorithm}

\begin{algorithm}[H]
\caption{Bottenbruch's Binary Search with $[l, r]$ bounds}
\begin{algorithmic}[1]
\Function{bsearch5}{$A, t$}
    \State $l \gets 0$
    \State $r \gets n - 1$  \Comment{$r$ is inclusive}
    \While{$l < r$} \Comment{Note: $<$ not $\leq$, even with inclusive bounds!}
        \State $m \gets l + \lceil(r - l) / 2\rceil$ \Comment{Ceiling, not floor!}
        \If{$A[m] > t$}
            \State $r \gets m - 1$  \Comment{Target in $[l,m-1]$}
        \Else
            \State $l \gets m$ \Comment{$l$ can stay at $m$ (no $+1$)}
        \EndIf
        \EndWhile
    \If{$l \leq r$ \textbf{and} $A[l] = t$} \Comment{if $A$ is not empty}
        \State \Return $l$  \Comment{Target found}
    \EndIf
    \State \Return $-1$  \Comment{Target not found}
\EndFunction
\end{algorithmic}
\end{algorithm}

\subsection{The Ceiling Midpoint: Avoiding Infinite Loops}

\begin{tcolorbox}[colback=red!5, colframe=red!75!black, title=Critical Detail]
Bottenbruch uses $m = l + \lceil(r - l) / 2\rceil$ instead of the usual floor. In integer arithmetic this is computed as $l + \lfloor(r - l + 1) / 2\rfloor$, which is the form used in the code listings.

\textbf{Why?} When $l = m$ in the update and $r - l = 1$:
\begin{itemize}[noitemsep]
    \item With floor: $m = l + \lfloor 1/2 \rfloor = l$, so if $A[m] \leq t$, we set $l = m = l$ (no progress!)
    \item With ceiling: $m = l + \lceil 1/2 \rceil = l + 1$, ensuring progress
\end{itemize}
\end{tcolorbox}

\subsection{Loop Invariant}

\begin{invariant}[bsearch5]
At the start of each iteration:
\begin{enumerate}[noitemsep]
    \item $0 \leq l \leq r \leq n - 1$
    \item If $t$ exists in $A$, then $t \in A[l..r]$
    \item All elements $A[0..l-1]$ are $< t$
    \item All elements $A[r+1..n-1]$ are $> t$
\end{enumerate}
\end{invariant}

\subsection{Post-condition}

\begin{postcondition}
When the loop terminates ($l = r$):
\begin{itemize}[noitemsep]
    \item $l = r$ is the \emph{rightmost} index such that $A[l] = t$ (if $t$ exists)
    \item This follows because the update $l \gets m$ keeps searching rightward when $A[m] \leq t$
    \item We must explicitly check $A[l] = t$ since the loop defers the equality test
\end{itemize}
\end{postcondition}

\subsection{Comparison with Bentley (Variant 2)}

\begin{center}
\begin{tabular}{lcc}
\toprule
\textbf{Aspect} & \textbf{Bentley (bsearch2)} & \textbf{Bottenbruch (bsearch5)} \\
\midrule
Bounds & $[l, r]$ (both inclusive) & $[l, r]$ (both inclusive) \\
Loop condition & $l \leq r$ & $l < r$ \\
Equality check & Inside loop & After loop \\
Midpoint & $l + \lfloor (r - l) / 2 \rfloor$ & $l + \lceil (r - l) / 2 \rceil$ \\
Update $l$ & $m + 1$ & $m$ \\
Comparisons per iter & 2 (avg) & 1 \\
\bottomrule
\end{tabular}
\end{center}

The advantage of Bottenbruch is \emph{one comparison per iteration}
inside the loop (only the $A[m] > t$ check), which can be faster in
practice despite potentially one extra iteration.

\begin{tcolorbox}[colback=blue!5, colframe=blue!75!black, title=Behavioral Note]
Unlike the 3-way comparison variants (which return an arbitrary
match), Bottenbruch's algorithm \textbf{always returns the rightmost}
occurrence of $t$ when duplicates exist \citep{wikipedia_bsearch}. This
is because when $A[m] = t$, we set $l = m$, continuing to search
rightward.
\end{tcolorbox}

\begin{tcolorbox}[invbox={Performance: Monobound Binary Search}]
The deferred-equality approach has been independently rediscovered and
optimized by \citet{scandum2020} as the \emph{monobound binary
  search}. By eliminating the equality check inside
the loop, the algorithm achieves 2--4$\times$ faster execution on
arrays smaller than 1 million elements due to better branch
prediction. The key insight: modern CPUs handle predictable branches
(always $<$ or $\geq$) more efficiently than unpredictable 3-way
branches.

\medskip
\textbf{Implementation note:} Van den Hoven's monobound uses
\emph{count-based} iteration (a size variable \texttt{top} that
tracks the remaining search space) rather than index-based $[l, r]$
bounds. Count-based implementations naturally avoid the ceiling
midpoint requirement because the size variable strictly decreases
($\mathit{top} \gets \mathit{top} - \mathit{mid}$) at every step,
guaranteeing progress without the $l = m$ stall that index-based
formulations must guard against. The BSD (Section~\ref{sec:bsd}) and
C++ STL (Section~\ref{sec:cplusplus}) implementations use the same
count-based pattern.
\end{tcolorbox}

\subsection{Python Implementation}

\begin{lstlisting}[style=pythonstyle, caption=Variant 5: Bottenbruch's algorithm (finds rightmost)]
def bsearch5(A: list[int], t: int) -> int:
  l, r = 0, len(A) - 1  # r inclusive
  while l < r:  # Note: < not <=, even with inclusive bounds
    m = l + (r - l + 1) // 2  # ceil: l + ceil((r-l)/2)
    if A[m] > t:
      r = m - 1         # Target in [l,m-1]
    else:
      l = m             # l can stay at m (no +1)
  if l <= r and A[l] == t:  # if A is not empty
    return l            # Target found
  return -1             # Target not found
\end{lstlisting}

\subsection{Dafny Implementation (Formally Verified)}

The Dafny version matches the Python implementation, returning $-1$ if not found:

\begin{lstlisting}[style=dafnystyle, caption=Variant 5 in Dafny (Bottenbruch)]
method bsearch5(A: array<int>, t: int) returns (idx: int)
  requires Sorted(A)
  requires A.Length <= MAX_INT32
  ensures -1 <= idx < A.Length
  ensures idx != -1 ==> A[idx] == t
  ensures idx == -1 ==> forall k :: 0 <= k < A.Length ==> A[k] != t
{
  if A.Length == 0 {
    return -1;  // Invariant requires l <= r; empty array gives r = -1
  }
  var l := 0;
  var r := A.Length - 1;  // r inclusive
  while l < r
    invariant 0 <= l <= r <= A.Length - 1
    invariant (exists k :: 0 <= k < A.Length && A[k] == t) 
              ==> (exists k :: l <= k <= r && A[k] == t)
    invariant forall k :: r < k < A.Length ==> A[k] > t
    invariant forall k :: 0 <= k < l ==> A[k] <= t
    decreases r - l
  {
    assert l + (r - l + 1) / 2 <= MAX_INT32;  // Overflow check
    var m := l + (r - l + 1) / 2;             // CEILING division
    if A[m] > t {
      r := m - 1;         // Target in [l,m-1]
    } else {              // l can stay at m (no +1)
      l := m;             // Target in [m,r]
    }  
  }
  if l <= r && A[l] == t {    // if A is not empty
    return l;             // Target found
  } 
  return -1;              // Target not found
}
\end{lstlisting}

\section{Approximate Matches and Derived Queries}
\label{sec:derived}

Binary search's utility extends beyond exact matches. The leftmost and
rightmost variants enable a family of \emph{approximate match} queries
on sorted arrays \citep{knuth1998, sedgewick2011}. All derived functions build upon
\texttt{bisect\_left} and \texttt{bisect\_right}, providing $O(\log
n)$ query time.

\subsection{find\_rank: Count Elements Less Than \texorpdfstring{$t$}{t}}

The \emph{rank} of a value $t$ is the number of elements strictly
less than $t$. This is exactly what \texttt{bisect\_left} returns:
\[
\texttt{find\_rank}(A, t) = \texttt{bisect\_left}(A, t) = |\{i : A[i] < t\}|
\]

\textbf{Return value}: An integer in $[0, n]$ representing the count of elements $< t$.

\textbf{Examples} for $A = [1, 3, 3, 5, 7]$:
\begin{itemize}[noitemsep]
    \item $\texttt{find\_rank}(A, 3) = 1$ (one element less than 3)
    \item $\texttt{find\_rank}(A, 4) = 3$ (three elements less than 4)
    \item $\texttt{find\_rank}(A, 0) = 0$ (no elements less than 0)
    \item $\texttt{find\_rank}(A, 10) = 5$ (all elements less than 10)
\end{itemize}

\subsection{find\_pred\_strict: Strict Predecessor (Largest Element
  \texorpdfstring{$< t$}{< t})}

The \emph{strict predecessor} finds the largest element strictly
less than $t$. If $\text{rank}(t) = r$, then the strict predecessor is
at index $r - 1$ (if $r > 0$).

\[
\texttt{find\_pred\_strict}(A, t) = 
\begin{cases}
\texttt{bisect\_left}(A, t) - 1 & \text{if } \texttt{bisect\_left}(A, t) > 0 \\
-1 & \text{otherwise (no element} < t\text{)}
\end{cases}
\]

\textbf{Return value}: Index of the largest element $< t$, or $-1$ if all elements are $\geq t$.

\textbf{Examples} for $A = [1, 3, 3, 5, 7]$:
\begin{itemize}[noitemsep]
    \item $\texttt{find\_pred\_strict}(A, 5) = 2$ (returns index of 3)
    \item $\texttt{find\_pred\_strict}(A, 3) = 0$ (returns index of 1)
    \item $\texttt{find\_pred\_strict}(A, 1) = -1$ (no element $< 1$)
\end{itemize}

\subsection{find\_floor: Floor / Non-Strict Predecessor (Largest
  Element \texorpdfstring{$\leq t$}{<= t})}

The \emph{floor} (or non-strict predecessor) finds the largest
element less than or equal to $t$. This uses \texttt{bisect\_right}
because we want to include elements equal to $t$.

\[
\texttt{find\_floor}(A, t) = 
\begin{cases}
\texttt{bisect\_right}(A, t) - 1 & \text{if } \texttt{bisect\_right}(A, t) > 0 \\
-1 & \text{otherwise (no element} \leq t\text{)}
\end{cases}
\]

\textbf{Return value}: Index of the largest element $\leq t$, or $-1$ if all elements are $> t$.

\textbf{Examples} for $A = [1, 3, 3, 5, 7]$:
\begin{itemize}[noitemsep]
    \item $\texttt{find\_floor}(A, 5) = 3$ (returns index of 5 itself)
    \item $\texttt{find\_floor}(A, 4) = 2$ (returns index of rightmost 3)
    \item $\texttt{find\_floor}(A, 0) = -1$ (no element $\leq 0$)
\end{itemize}

\textbf{Key difference from find\_pred\_strict}: When $t$ exists in
the array, \texttt{find\_floor} returns $t$'s index while
\texttt{find\_pred\_strict} returns the element before $t$.

\subsection{find\_succ\_strict: Strict Successor (Smallest Element
  \texorpdfstring{$> t$}{> t})}

The \emph{strict successor} finds the smallest element strictly
greater than $t$. This is the element at the insertion point returned
by \texttt{bisect\_right}.

\[
\texttt{find\_succ\_strict}(A, t) = 
\begin{cases}
\texttt{bisect\_right}(A, t) & \text{if } \texttt{bisect\_right}(A, t) < n \\
-1 & \text{otherwise (no element} > t\text{)}
\end{cases}
\]

\textbf{Return value}: Index of the smallest element $> t$, or $-1$ if all elements are $\leq t$.

\textbf{Examples} for $A = [1, 3, 3, 5, 7]$:
\begin{itemize}[noitemsep]
    \item $\texttt{find\_succ\_strict}(A, 3) = 3$ (returns index of 5)
    \item $\texttt{find\_succ\_strict}(A, 4) = 3$ (returns index of 5)
    \item $\texttt{find\_succ\_strict}(A, 7) = -1$ (no element $> 7$)
\end{itemize}

\subsection{find\_ceil: Ceiling / Non-Strict Successor (Smallest
  Element \texorpdfstring{$\geq t$}{>= t})}

The \emph{ceiling} (or non-strict successor) finds the smallest
element greater than or equal to $t$. This uses \texttt{bisect\_left}
because we want to include elements equal to $t$.

\[
\texttt{find\_ceil}(A, t) = 
\begin{cases}
\texttt{bisect\_left}(A, t) & \text{if } \texttt{bisect\_left}(A, t) < n \\
-1 & \text{otherwise (no element} \geq t\text{)}
\end{cases}
\]

\textbf{Return value}: Index of the smallest element $\geq t$, or $-1$ if all elements are $< t$.

\textbf{Examples} for $A = [1, 3, 3, 5, 7]$:
\begin{itemize}[noitemsep]
    \item $\texttt{find\_ceil}(A, 3) = 1$ (returns index of leftmost 3)
    \item $\texttt{find\_ceil}(A, 4) = 3$ (returns index of 5)
    \item $\texttt{find\_ceil}(A, 8) = -1$ (no element $\geq 8$)
\end{itemize}

\textbf{Key difference from find\_succ\_strict}: When $t$ exists in
the array, \texttt{find\_ceil} returns $t$'s index while
\texttt{find\_succ\_strict} returns the element after $t$.

\subsection{find\_range: Count Elements in \texorpdfstring{$[a,
      b]$}{[a, b]}}

The \emph{range query} counts elements in the closed interval $[a,
  b]$. This combines \texttt{bisect\_left} and \texttt{bisect\_right}:

\[
\texttt{find\_range}(A, a, b) = \max\bigl(0,\; \texttt{bisect\_right}(A, b) - \texttt{bisect\_left}(A, a)\bigr) = |\{i : a \leq A[i] \leq b\}|
\]

\textbf{Return value}: Count of elements in $[a, b]$, always $\geq 0$. The clamp to zero matters only in the degenerate case $a > b$ (an empty interval), where the raw difference \texttt{bisect\_right}$(A, b) - \texttt{bisect\_left}(A, a)$ can be negative; for $a \leq b$ the difference is already non-negative.

\textbf{Examples} for $A = [1, 3, 3, 5, 7]$:
\begin{itemize}[noitemsep]
    \item $\texttt{find\_range}(A, 2, 5) = 3$ (elements 3, 3, 5)
    \item $\texttt{find\_range}(A, 3, 3) = 2$ (counts duplicates of 3)
    \item $\texttt{find\_range}(A, 10, 20) = 0$ (no elements in range)
\end{itemize}

\textbf{Special case---counting duplicates}: To count occurrences of value $t$:
\[
\text{count\_duplicates}(t) = \texttt{find\_range}(A, t, t)
\]

\subsection{Summary of Derived Functions}

\begin{center}
\begin{tabular}{llll}
\toprule
\textbf{Function} & \textbf{Query} & \textbf{Based On} & \textbf{Returns $-1$ When} \\
\midrule
\texttt{find\_rank} & count $< t$ & \texttt{bisect\_left} & never (returns 0) \\
\texttt{find\_pred\_strict} & max $< t$ & \texttt{bisect\_left} & all $\geq t$ \\
\texttt{find\_floor} & max $\leq t$ & \texttt{bisect\_right} & all $> t$ \\
\texttt{find\_succ\_strict} & min $> t$ & \texttt{bisect\_right} & all $\leq t$ \\
\texttt{find\_ceil} & min $\geq t$ & \texttt{bisect\_left} & all $< t$ \\
\texttt{find\_range} & count in $[a,b]$ & both & never (returns 0) \\
\bottomrule
\end{tabular}
\end{center}

\subsection{Nearest Neighbor}
The nearest neighbor is either the floor or ceiling, whichever is closer in value:
\[
\text{nearest}(t) = \arg\min_{x \in \{\text{floor}(t), \text{ceil}(t)\}} |x - t|
\]

\subsection{Python Implementations of Derived Functions}

The following functions build on \texttt{bisect\_left} and \texttt{bisect\_right}:

\begin{lstlisting}[style=pythonstyle, caption=Derived query functions]
def find_rank(A: list[int], t: int) -> int:
  """Count elements strictly less than t."""
  return bisect_left(A, t)

def find_pred_strict(A: list[int], t: int) -> int:
  """Find largest index i where A[i] < t, or -1."""
  i = bisect_left(A, t)
  return i - 1 if i > 0 else -1

def find_floor(A: list[int], t: int) -> int:
  """Find largest index i where A[i] <= t, or -1."""
  i = bisect_right(A, t)
  return i - 1 if i > 0 else -1

def find_succ_strict(A: list[int], t: int) -> int:
  """Find smallest index i where A[i] > t, or -1."""
  i = bisect_right(A, t)
  return i if i < len(A) else -1

def find_ceil(A: list[int], t: int) -> int:
  """Find smallest index i where A[i] >= t, or -1."""
  i = bisect_left(A, t)
  return i if i < len(A) else -1

def find_range(A: list[int], a: int, b: int) -> int:
  """Count elements in [a, b]; 0 if a > b."""
  return max(0, bisect_right(A, b) - bisect_left(A, a))
\end{lstlisting}

\subsection{Dafny Implementations (Formally Verified)}

The following Dafny methods are formally verified to satisfy their
specifications. These implementations call \texttt{bisect\_left} and
\texttt{bisect\_right} (Section~\ref{sec:bisect-dafny}), whose
cardinality proofs rely on the \texttt{RangeSet} and
\texttt{SetDifferenceCardinality} auxiliary lemmas described in
Section~\ref{sec:dafny-lemmas} and detailed in
Appendix~\ref{sec:dafny-auxiliary}.

\begin{lstlisting}[style=dafnystyle, caption=Derived query functions in Dafny (formally verified)]
method find_rank(A: array<int>, t: int) returns (count: int)
  requires Sorted(A)
  requires A.Length <= MAX_INT32
  ensures 0 <= count <= A.Length
  ensures count == |set k | 0 <= k < A.Length && A[k] < t|
{
  count := bisect_left(A, t);  // Inherits cardinality proof from bisect_left
}

method find_pred_strict(A: array<int>, t: int) returns (idx: int)
  requires Sorted(A)
  requires A.Length <= MAX_INT32
  ensures -1 <= idx < A.Length
  ensures idx != -1 ==> A[idx] < t
  ensures idx != -1 ==> (idx == A.Length - 1 || A[idx + 1] >= t)
  ensures idx == -1 ==> forall k :: 0 <= k < A.Length ==> A[k] >= t
{
  var i := bisect_left(A, t);
  if i > 0 {
    return i - 1;
  }
  return -1;
}

method find_floor(A: array<int>, t: int) returns (idx: int)
  requires Sorted(A)
  requires A.Length <= MAX_INT32
  ensures -1 <= idx < A.Length
  ensures idx != -1 ==> A[idx] <= t
  ensures idx != -1 ==> (idx == A.Length - 1 || A[idx + 1] > t)
  ensures idx == -1 ==> forall k :: 0 <= k < A.Length ==> A[k] > t
{
  var i := bisect_right(A, t);
  if i > 0 {
    return i - 1;
  }
  return -1;
}

method find_succ_strict(A: array<int>, t: int) returns (idx: int)
  requires Sorted(A)
  requires A.Length <= MAX_INT32
  ensures -1 <= idx < A.Length
  ensures idx != -1 ==> A[idx] > t
  ensures idx != -1 ==> (idx == 0 || A[idx - 1] <= t)
  ensures idx == -1 ==> forall k :: 0 <= k < A.Length ==> A[k] <= t
{
  var i := bisect_right(A, t);
  if i < A.Length {
    return i;
  }
  return -1;
}

method find_ceil(A: array<int>, t: int) returns (idx: int)
  requires Sorted(A)
  requires A.Length <= MAX_INT32
  ensures -1 <= idx < A.Length
  ensures idx != -1 ==> A[idx] >= t
  ensures idx != -1 ==> (idx == 0 || A[idx - 1] < t)
  ensures idx == -1 ==> forall k :: 0 <= k < A.Length ==> A[k] < t
{
  var i := bisect_left(A, t);
  if i < A.Length {
    return i;
  }
  return -1;
}

method find_range(A: array<int>, a: int, b: int) returns (count: int)
  requires Sorted(A)
  requires A.Length <= MAX_INT32
  ensures 0 <= count <= A.Length
  ensures count == |set k | 0 <= k < A.Length && a <= A[k] <= b|
{
  var left := bisect_left(A, a);
  var right := bisect_right(A, b);
  
  // Define sets for proof
  var S_range := set k | 0 <= k < A.Length && a <= A[k] <= b;
  var S_left  := set k | 0 <= k < A.Length && A[k] < a;
  var S_right := set k | 0 <= k < A.Length && A[k] <= b;

  // Handle edge case: if left > right, the range is empty
  // This can happen when a > b, or when no elements fall in [a, b]
  if left > right {
      assert forall k :: k in S_range ==> false;
      assert S_range == {};
      return 0;
  }

  count := right - left;

  // Proof block: Show count == |S_range|
  // Step 1: Prove S_left <= S_right (subset relationship)
  assert forall k :: k in S_left ==> k < left;
  assert forall k :: 0 <= k < right ==> k in S_right;
  assert forall k :: k in S_left ==> k in S_right;
  assert S_left <= S_right;
  
  // Step 2: Prove S_range == S_right - S_left
  // An element is in [a, b] iff it's <= b but not < a
  assert S_range == S_right - S_left;
  
  // Step 3: Use SetDifferenceCardinality lemma
  SetDifferenceCardinality(S_right, S_left);
  // Now: |S_range| == |S_right| - |S_left|
  
  // Step 4: Connect set cardinalities to indices using RangeSet
  RangeSetProperties(right);
  RangeSetProperties(left);
  assert S_right == RangeSet(right);  // |S_right| == right
  assert S_left  == RangeSet(left);   // |S_left| == left
  // Therefore: |S_range| == right - left == count
}
\end{lstlisting}

\begin{tcolorbox}[invbox={Practical Note}]
C++'s \texttt{<algorithm>} header provides \texttt{std::lower\_bound}, \texttt{std::upper\_bound}, and \texttt{std::equal\_range} which implement these patterns. Python's \texttt{bisect} module provides \texttt{bisect\_left} and \texttt{bisect\_right}.
\end{tcolorbox}

\section{Python's bisect Module}
\label{sec:bisect}

Python's standard library \citep{cpython_bisect} provides
\texttt{bisect\_left} and \texttt{bisect\_right}, which find insertion
points rather than exact matches. The following implementations use
our standard variable names (A, t, l, r, m), which map to Python's
stdlib names as: $a \to A$, $x \to t$, $lo \to l$, $hi \to r$, $mid
\to m$. The original functions also accept optional \texttt{lo} and
\texttt{hi} parameters to restrict the search to a subrange $[lo,
  hi)$; we show the default case for clarity.

\subsection{bisect\_left}

\begin{algorithm}[H]
\caption{bisect\_left: Find leftmost insertion point}
\begin{algorithmic}[1]
\Function{bisect\_left}{$A, t$}
    \State $l \gets 0$, $r \gets n$  \Comment{$r$ exclusive; insertion point in $[l,r]$}
    \While{$l < r$}
        \State $m \gets l + \lfloor(r - l) / 2\rfloor$
        \If{$A[m] < t$}
            \State $l \gets m + 1$  \Comment{Insertion point in $[m+1,r]$}
        \Else
            \State $r \gets m$  \Comment{Insertion point in $[l,m]$}
        \EndIf
    \EndWhile
    \State \Return $l$  \Comment{Insertion point ($l = r$)}
\EndFunction
\end{algorithmic}
\end{algorithm}

\textbf{Semantics}: Returns index $i$ such that all elements in
$A[0..i)$ are $< t$, and all elements in $A[i..n)$ are $\geq t$.

\subsection{bisect\_right}

\begin{algorithm}[H]
\caption{bisect\_right: Find rightmost insertion point}
\begin{algorithmic}[1]
\Function{bisect\_right}{$A, t$}
    \State $l \gets 0$, $r \gets n$  \Comment{$r$ exclusive; insertion point in $[l,r]$}
    \While{$l < r$}
        \State $m \gets l + \lfloor(r - l) / 2\rfloor$
        \If{$t < A[m]$}
            \State $r \gets m$  \Comment{Insertion point in $[l,m]$}
        \Else
            \State $l \gets m + 1$  \Comment{Insertion point in $[m+1,r]$}
        \EndIf
    \EndWhile
    \State \Return $l$  \Comment{Insertion point ($l = r$)}
\EndFunction
\end{algorithmic}
\end{algorithm}

\textbf{Semantics}: Returns index $i$ such that all elements in
$A[0..i)$ are $\leq t$, and all elements in $A[i..n)$ are $> t$.

\paragraph{A note on the bracket convention.}
The in-loop comments for the \emph{search} variants
(Sections~\ref{sec:variant1}--\ref{sec:variant5}) use half-open
intervals over array \emph{indices}, where $r$ is excluded because
$A[r]$ is outside the window. The comments above instead bracket the
\emph{insertion point}, a position in $[0, n]$; the interval is closed
at both ends because the current $r$ is always itself a feasible
insertion position.

\subsection{Python Implementations}

\begin{lstlisting}[style=pythonstyle, caption=bisect\_left: Find leftmost insertion point]
def bisect_left(A: list[int], t: int) -> int:
  """Return insertion point for t to maintain sorted order.
  If t is already present, insertion point is before (left of) any existing entries.
  Stdlib names: a -> A, x -> t, lo -> l, hi -> r, mid -> m"""
  l, r = 0, len(A)    # r exclusive
  while l < r:
    m = l + (r - l) // 2    # floor: l + floor((r-l)/2)
    if A[m] < t:
      l = m + 1
    else:
      r = m
  return l    # Always returns a valid insertion point [0, len(A)]
\end{lstlisting}

\begin{lstlisting}[style=pythonstyle, caption=bisect\_right: Find rightmost insertion point]
def bisect_right(A: list[int], t: int) -> int:
  """Return insertion point for t to maintain sorted order.
  If t is already present, insertion point is after (right of) any existing entries.
  Stdlib names: a -> A, x -> t, lo -> l, hi -> r, mid -> m"""
  l, r = 0, len(A)   # r exclusive
  while l < r:
    m = l + (r - l) // 2    # floor: l + floor((r-l)/2)
    if A[m] > t:     
      r = m
    else:
      l = m + 1
  return l    # Always returns a valid insertion point [0, len(A)]
\end{lstlisting}

\subsection{Using bisect for Exact Search}

\begin{lstlisting}[style=pythonstyle, caption=Exact search using bisect\_left]
def binary_search(A, t):
  i = bisect_left(A, t)
  if i < len(A) and A[i] == t:
    return i
  return -1
\end{lstlisting}

\subsection{Dafny Implementations (Formally Verified)}
\label{sec:bisect-dafny}

The Dafny implementations of \texttt{bisect\_left} and
\texttt{bisect\_right} include cardinality proofs connecting the
returned index to the size of the corresponding element set. These
proofs use the \texttt{RangeSet} and \texttt{RangeSetProperties}
auxiliary lemmas described in Section~\ref{sec:dafny-lemmas} and
detailed in Appendix~\ref{sec:dafny-auxiliary}.

\begin{lstlisting}[style=dafnystyle, caption=bisect\_left in Dafny with cardinality proof]
method bisect_left(A: array<int>, t: int) returns (idx: int)
  requires Sorted(A)
  requires A.Length <= MAX_INT32
  ensures 0 <= idx <= A.Length
  ensures forall k :: 0 <= k < idx ==> A[k] < t
  ensures forall k :: idx <= k < A.Length ==> A[k] >= t
  ensures idx == |set k | 0 <= k < A.Length && A[k] < t|  // Cardinality
{
  var l := 0;
  var r := A.Length;   // r exclusive
  while l < r
    invariant 0 <= l <= r <= A.Length
    invariant forall k :: 0 <= k < l ==> A[k] < t
    invariant forall k :: r <= k < A.Length ==> A[k] >= t
    decreases r - l
  {
    assert l + (r - l) / 2 <= MAX_INT32;  // Overflow check
    var m := l + (r - l) / 2;
    if A[m] < t {
      l := m + 1;      // Target in [m+1,r)     
    } else {
      r := m;          // Target in [l,m)
    }
  }

  // Proof block: Connect the index l to set cardinality
  // The set of indices where A[k] < t is exactly {0, 1, ..., l-1}
  var s := set k | 0 <= k < A.Length && A[k] < t;
  assert forall k :: 0 <= k < l ==> k in s;      // All k < l are in s
  assert forall k :: l <= k < A.Length ==> k !in s;  // All k >= l are not in s
  
  // Use RangeSet to prove |s| == l without trigger warnings
  RangeSetProperties(l);
  assert s == RangeSet(l);  // s is exactly {0, 1, ..., l-1}
  
  return l;
}
\end{lstlisting}

\begin{lstlisting}[style=dafnystyle, caption=bisect\_right in Dafny with cardinality proof]
method bisect_right(A: array<int>, t: int) returns (idx: int)
  requires Sorted(A)
  requires A.Length <= MAX_INT32
  ensures 0 <= idx <= A.Length
  ensures forall k :: 0 <= k < idx ==> A[k] <= t
  ensures forall k :: idx <= k < A.Length ==> A[k] > t
  ensures idx == |set k | 0 <= k < A.Length && A[k] <= t|  // Cardinality
{
  var l := 0;
  var r := A.Length;
  while l < r
    invariant 0 <= l <= r <= A.Length
    invariant forall k :: 0 <= k < l ==> A[k] <= t
    invariant forall k :: r <= k < A.Length ==> A[k] > t
    decreases r - l
  {
    var m := l + (r - l) / 2;
    if A[m] > t {
      r := m;
    } else {
      l := m + 1;
    }
  }
  // Proof: Connect index to set cardinality using RangeSet lemma
  var s := set k | 0 <= k < A.Length && A[k] <= t;
  RangeSetProperties(l);
  assert s == RangeSet(l);  // |s| == l
  return l;
}
\end{lstlisting}

\section{Standard Library Implementations}
\label{sec:stdlib}

Multiple standard libraries provide binary search implementations with
varying conventions. We survey implementations from C (BSD and glibc),
Java (OpenJDK), and C++ (STL).

\subsection{BSD's Implementation (Apple)}
\label{sec:bsd}

\noindent\textit{Category: 3-Way Comparison with limit-based iteration}

\medskip
The C standard library provides \texttt{bsearch()}, using a comparator
callback returning a negative, zero, or positive integer.
The following implementations use our standard variable names
($A$, $t$, $n$, $l$, $m$), which map to BSD's original names as:
$\texttt{key} \to t$, $\texttt{base} \to A$, $\texttt{nmemb} \to n$,
$\texttt{lim}$ for remaining count, $\texttt{p} \to m$.

\medskip
The BSD \texttt{bsearch} \citep{bsd_bsearch}, shipped in Apple's libc,
uses an interesting approach with a ``limit'' variable that tracks the
remaining search space size:

\subsubsection{Algorithm}

\begin{algorithm}[H]
\caption{BSD bsearch (Apple libc)}
\begin{algorithmic}[1]
\Function{bsearch}{$key, base, nmemb$}
    \For{$lim \gets nmemb$; $lim \neq 0$; $lim \gets lim \gg 1$}
        \State $p \gets base + (lim \gg 1)$
        \State $cmp \gets compare(key, p)$
        \If{$cmp = 0$}
            \State \Return $p$
        \EndIf
        \If{$cmp > 0$} \Comment{key $> p$: move right}
            \State $base \gets p + 1$
            \State $lim \gets lim - 1$
        \EndIf
    \EndFor
    \State \Return NULL
\EndFunction
\end{algorithmic}
\end{algorithm}

\subsubsection{Loop Invariant}

\begin{invariant}[BSD bsearch]
At the start of each iteration:
\begin{enumerate}[noitemsep]
    \item \texttt{lim} is the number of elements remaining to search
    \item If key exists, it is within \texttt{base[0..lim-1]}
    \item All elements before current \texttt{base} are $<$ key
\end{enumerate}
\end{invariant}

\subsubsection{Key Insight}

The $lim \gets lim - 1$ before halving when moving right accounts for
the odd/even asymmetry elegantly. When we move right past the
midpoint, we've eliminated $\lfloor lim/2 \rfloor + 1$ elements,
leaving $lim - \lfloor lim/2 \rfloor - 1$ elements.

\subsubsection{Python Implementation}

\begin{lstlisting}[style=pythonstyle, caption=BSD-style bsearch with limit variable]
def bsearch_bsd(A: list[int], t: int) -> int:
  """BSD-style binary search using limit-based iteration.
  C names: key -> t, base -> l (as index), nmemb -> n, lim -> remaining, p -> m
  Returns index of any matching element, or -1 if not found."""
  l = 0                  # base: starting index
  lim = len(A)           # nmemb: remaining elements
  while lim != 0:
    m = l + (lim >> 1)   # p: midpoint (lim >> 1 = floor(lim/2))
    if A[m] == t:
      return m
    if A[m] < t:         # key > A[m]: move right
      l = m + 1          # base = p + 1
      lim -= 1           # Account for skipping m
    lim >>= 1            # Halve remaining search space
  return -1
\end{lstlisting}

\subsubsection{Dafny Implementation (Formally Verified)}

The Dafny version matches the Python implementation, returning $-1$ if not found:

\begin{lstlisting}[style=dafnystyle, caption=BSD-style bsearch in Dafny]
method bsearch_bsd(A: array<int>, t: int) returns (idx: int)
  requires Sorted(A)
  requires A.Length <= MAX_INT32
  ensures -1 <= idx < A.Length
  ensures idx != -1 ==> A[idx] == t
  ensures idx == -1 ==> forall k :: 0 <= k < A.Length ==> A[k] != t
{
  var l := 0;           // base: starting index
  var lim := A.Length;  // nmemb: remaining elements
  while lim != 0
    invariant 0 <= l <= A.Length
    invariant 0 <= lim <= A.Length
    invariant l + lim <= A.Length
    invariant forall k :: 0 <= k < l ==> A[k] < t
    invariant forall k :: l + lim <= k < A.Length ==> A[k] > t
    decreases lim
  {
    var m := l + lim / 2;  // p: midpoint
    if A[m] == t {         // Target found
      return m;
    }
    if A[m] < t {          // key > A[m]: move right
      l := m + 1;          // base = p + 1
      lim := lim - 1;      // Account for skipping m
    }
    lim := lim / 2;        // Halve remaining search space
  }
  return -1;
}
\end{lstlisting}

\subsection{glibc's Implementation (GNU C Library)}

\noindent\textit{Category: 3-Way Comparison with $[l, r)$ bounds}

\medskip
The GNU C Library (glibc), the libc used by GCC toolchains on Linux,
implements \texttt{bsearch} \citep{gcc_bsearch} following the
standard $[l, r)$ convention, identical to our Variant~1. (GCC's own
libiberty ships a separate, BSD-derived limit-based
\texttt{bsearch}; the implementation analyzed here is glibc's. Our
accompanying code retains the identifier \texttt{bsearch\_gcc} for
this variant.) The following implementations use our standard
variable names, which map to glibc's original names as:
$\texttt{key} \to t$, $\texttt{base} \to A$, $\texttt{nmemb} \to n$,
$\texttt{l} \to l$, $\texttt{u} \to r$, $\texttt{idx} \to m$.

\subsubsection{Algorithm}

\begin{algorithm}[H]
\caption{glibc bsearch}
\begin{algorithmic}[1]
\Function{bsearch}{$key, base, nmemb$}
    \State $l \gets 0$, $u \gets nmemb$
    \While{$l < u$}
        \State $idx \gets (l + u) / 2$
        \State $cmp \gets compare(key, base[idx])$
        \If{$cmp < 0$}
            \State $u \gets idx$
        \ElsIf{$cmp > 0$}
            \State $l \gets idx + 1$
        \Else
            \State \Return $base[idx]$
        \EndIf
    \EndWhile
    \State \Return NULL
\EndFunction
\end{algorithmic}
\end{algorithm}

\subsubsection{Loop Invariant}

\begin{invariant}[glibc bsearch]
At the start of each iteration:
\begin{enumerate}[noitemsep]
    \item $0 \leq l \leq u \leq nmemb$
    \item If key exists, it is within \texttt{base[l..u-1]}
    \item All elements in \texttt{base[0..l-1]} are $<$ key
    \item All elements in \texttt{base[u..nmemb-1]} are $>$ key
\end{enumerate}
\end{invariant}

\subsubsection{Equivalence to Variant 1}

This is essentially identical to our bsearch1, using $[l, u)$
  half-open interval convention. There are two minor differences: the
  use of a comparator function returning $\{-1, 0, +1\}$ instead of
  direct comparison operators, and the midpoint form. glibc's original
  computes $(l + u)/2$, which cannot overflow in practice because the
  indices are unsigned \texttt{size\_t} values bounded by the array
  length; our transcription uses the overflow-safe $l + \lfloor(r -
  l)/2\rfloor$ idiom for consistency with the rest of the paper.

\subsubsection{Python Implementation}

\begin{lstlisting}[style=pythonstyle, caption=glibc-style bsearch with half-open interval]
def bsearch_gcc(A: list[int], t: int) -> int:
  """glibc-style binary search with [l, r) bounds.
  C names: key -> t, base -> A, nmemb -> n, l -> l, u -> r, idx -> m
  Equivalent to Variant 1 (bsearch1).
  Returns index of any matching element, or -1 if not found."""
  l, r = 0, len(A)          # u -> r
  while l < r:
    m = l + (r - l) // 2    # floor: overflow-safe midpoint
    if A[m] < t:
      l = m + 1
    elif A[m] > t:
      r = m
    else:
      return m
  return -1
\end{lstlisting}

\subsubsection{Dafny Implementation (Formally Verified)}

The Dafny version matches the Python implementation, returning $-1$ if not found:

\begin{lstlisting}[style=dafnystyle, caption=glibc-style bsearch in Dafny]
method bsearch_gcc(A: array<int>, t: int) returns (idx: int)
  requires Sorted(A)
  requires A.Length <= MAX_INT32
  ensures -1 <= idx < A.Length
  ensures idx != -1 ==> A[idx] == t
  ensures idx == -1 ==> forall k :: 0 <= k < A.Length ==> A[k] != t
{
  var l := 0;
  var r := A.Length;
  while l < r
    invariant 0 <= l <= r <= A.Length
    invariant forall k :: 0 <= k < l ==> A[k] < t
    invariant forall k :: r <= k < A.Length ==> A[k] > t
    decreases r - l
  {
    assert l + (r - l) / 2 <= MAX_INT32;  // Overflow check
    var m := l + (r - l) / 2;
    if A[m] < t {
      l := m + 1;
    } else if A[m] > t {
      r := m;
    } else {
      return m;
    }
  }
  return -1;
}
\end{lstlisting}

\subsection{Java's Implementation (OpenJDK)}

\noindent\textit{Category: 3-Way Comparison with $[l, r]$ bounds}

\medskip
Java's \texttt{Arrays.binarySearch} \citep{openjdk_bsearch} is
essentially Bentley's variant (bsearch2), with two distinguishing
features:

\begin{enumerate}[noitemsep]
  \item \textbf{Unsigned right shift for midpoint}: Java uses
    \texttt{(low + high)\urs 1}, which performs unsigned division even
    when \texttt{low + high} overflows a signed 32-bit integer. This is
    overflow-safe without needing the \texttt{l + (r - l) / 2} idiom.
  \item \textbf{Informative not-found return}: Instead of $-1$, Java
    returns $-(l + 1)$, encoding both ``not found'' and the insertion
    point. A non-negative return indicates the index of a match; a
    negative return can be decoded as \texttt{-result - 1} to obtain
    the insertion point.
\end{enumerate}

The following implementations use our standard variable names, which
map to Java's original names as: $\texttt{a} \to A$, $\texttt{key}
\to t$, $\texttt{low} \to l$, $\texttt{high} \to r$, $\texttt{mid}
\to m$. Java's \texttt{fromIndex} and \texttt{toIndex} parameters
correspond to $0$ and $n$.

\subsubsection{Algorithm}

\begin{algorithm}[H]
\caption{Java's Arrays.binarySearch (OpenJDK 11)}
\begin{algorithmic}[1]
\Function{binarySearch}{$A, t$}
    \State $l \gets 0$, $r \gets n - 1$  \Comment{low = fromIndex, high = toIndex $- 1$}
    \While{$l \leq r$}
        \State $m \gets (l + r) \ggg 1$  \Comment{Unsigned right shift; we use $l + \lfloor(r - l)/2\rfloor$}
        \If{$A[m] < t$}
            \State $l \gets m + 1$
        \ElsIf{$A[m] > t$}
            \State $r \gets m - 1$
        \Else
            \State \Return $m$  \Comment{Key found}
        \EndIf
    \EndWhile
    \State \Return $-(l + 1)$  \Comment{Not found; $l$ is the insertion point}
\EndFunction
\end{algorithmic}
\end{algorithm}

\subsubsection{Loop Invariant}

\begin{invariant}[Java bsearch]
At the start of each iteration:
\begin{enumerate}[noitemsep]
    \item $0 \leq l$ and $r \leq n - 1$ (with possible $l > r$ indicating empty interval)
    \item If key exists, it is within $A[l..r]$
    \item All elements in $A[0..l-1]$ are $<$ key
    \item All elements in $A[r+1..n-1]$ are $>$ key
\end{enumerate}
\end{invariant}

\subsubsection{Insertion Point Property}

When the loop exits without finding a match, $l = r + 1$ and the
invariants guarantee that all $A[0..l-1] < t$ and all $A[l..n-1] > t$.
Thus $l$ is exactly the \texttt{bisect\_left} insertion point: the
smallest index $i$ such that $A[i] \geq t$ (or $n$ if all elements are
less than $t$). The encoding $-(l + 1)$ is both negative (signaling
``not found'') and invertible: the caller recovers the insertion point
via $\texttt{-result} - 1$.

This makes Java's variant strictly more informative than the other
3-way comparison implementations (BSD, glibc), which return only a
boolean not-found indicator.

\subsubsection{Equivalence to Variant 2}

The search logic is identical to Bentley's bsearch2, using $[l, r]$
inclusive interval convention. The differences are the midpoint
calculation method (unsigned right shift vs.\ safe subtraction) and
the not-found return value encoding.

\subsubsection{Python Implementation}

\begin{lstlisting}[style=pythonstyle, caption=Java-style bsearch (OpenJDK)]
def bsearch_java(A: list[int], t: int) -> int:
  """Java's Arrays.binarySearch with [l, r] bounds.
  Returns >= 0: index of match. < 0: -(insertion_point + 1).
  Decode: insertion_point = -(result + 1) = bisect_left(A, t)."""
  l, r = 0, len(A) - 1         # high = toIndex - 1
  while l <= r:
    m = l + (r - l) // 2       # Java: (low + high) >>> 1
    if A[m] < t:
      l = m + 1
    elif A[m] > t:
      r = m - 1
    else:
      return m                  # key found
  return -(l + 1)               # not found; l is bisect_left position
\end{lstlisting}

\subsubsection{Dafny Implementation (Formally Verified)}

The Dafny version matches the Python implementation: a non-negative
return is the index of a match, and a negative return $-(l+1)$ encodes
the \texttt{bisect\_left} insertion point.

\begin{lstlisting}[style=dafnystyle, caption=Java-style bsearch in Dafny]
method bsearch_java(A: array<int>, t: int) returns (idx: int)
  requires Sorted(A)
  requires A.Length <= MAX_INT32
  // Found: non-negative index into a matching element
  ensures idx >= 0 ==> idx < A.Length && A[idx] == t
  // Not found: negative, encoding the bisect_left insertion point
  ensures idx < 0 ==> 0 <= -idx - 1 <= A.Length
  ensures idx < 0 ==> forall k :: 0 <= k < -idx - 1 ==> A[k] < t
  ensures idx < 0 ==> forall k :: -idx - 1 <= k < A.Length ==> A[k] > t
{
  var l := 0;
  var r := A.Length - 1;
  while l <= r
    invariant 0 <= l <= A.Length
    invariant -1 <= r < A.Length
    invariant l <= r + 1
    invariant forall k :: 0 <= k < l ==> A[k] < t
    invariant forall k :: r < k < A.Length ==> A[k] > t
    decreases r - l
  {
    assert l + (r - l) / 2 <= MAX_INT32;
    var m := l + (r - l) / 2;
    if A[m] < t {
      l := m + 1;
    } else if A[m] > t {
      r := m - 1;
    } else {
      return m;
    }
  }
  // l is the bisect_left insertion point; encode as -(l + 1)
  return -(l + 1);
}
\end{lstlisting}

\subsection{C++ STL Implementation}
\label{sec:cplusplus}

\noindent\textit{Category: 2-Way Comparison with count-based iteration}

\medskip
The C++ Standard Template Library \citep{cppreference_lower_bound}
decomposes binary search into two layers:
\texttt{std::lower\_bound} (the core algorithm) and
\texttt{std::binary\_search} (a thin wrapper).

Unlike all other standard library implementations surveyed here, the
C++ approach uses \textbf{2-way comparison} (only \texttt{<}) with
\textbf{count-based iteration}. This makes it structurally similar to
the BSD limit-based approach but applied to the deferred-equality
pattern. It finds the \textbf{leftmost} match when duplicates exist.

The following implementations use our standard variable names, which
map to C++'s original names as: $\texttt{first} \to l$,
$\texttt{count}$ for remaining elements, $\texttt{step}$ for
$\lfloor\text{count}/2\rfloor$, $\texttt{it} \to m$, $\texttt{value}
\to t$.

\subsubsection{Algorithm}

\begin{algorithm}[H]
\caption{C++ STL binary\_search via lower\_bound}
\begin{algorithmic}[1]
\Function{binary\_search}{$A, t$}
    \State \Comment{Phase 1: lower\_bound}
    \State $l \gets 0$, $count \gets n$
    \While{$count > 0$}
        \State $step \gets \lfloor count / 2 \rfloor$
        \State $m \gets l + step$
        \If{$A[m] < t$}
            \State $l \gets m + 1$
            \State $count \gets count - step - 1$
        \Else
            \State $count \gets step$
        \EndIf
    \EndWhile
    \State \Comment{Phase 2: equality check}
    \If{$l < n$ \textbf{and} $A[l] = t$}
        \State \Return $l$
    \EndIf
    \State \Return $-1$  \Comment{C++ original returns \texttt{bool}}
\EndFunction
\end{algorithmic}
\end{algorithm}

\subsubsection{Loop Invariant}

\begin{invariant}[C++ lower\_bound]
At the start of each iteration:
\begin{enumerate}[noitemsep]
    \item The search range is $A[l..l+count-1]$
    \item $0 \leq l \leq l + count \leq n$
    \item All elements in $A[0..l-1]$ are $< t$
    \item All elements in $A[l+count..n-1]$ are $\geq t$
\end{enumerate}
\end{invariant}

\subsubsection{Relationship to Other Variants}

\begin{center}
\begin{tabular}{lcc}
\toprule
\textbf{Aspect} & \textbf{BSD bsearch} & \textbf{C++ lower\_bound} \\
\midrule
Comparison & 3-way & 2-way \\
Iteration & count-based (\texttt{lim}) & count-based (\texttt{count}) \\
Equality check & Inside loop & After loop \\
Finds & Any match & Leftmost match \\
Equivalent to & bsearch1 & bsearch3 / bisect\_left \\
\bottomrule
\end{tabular}
\end{center}

\subsubsection{Python Implementation}

\begin{lstlisting}[style=pythonstyle, caption=C++ STL binary\_search via lower\_bound]
def bsearch_cplusplus(A: list[int], t: int) -> int:
  """C++ STL binary_search using lower_bound.
  C++ names: first -> l, count -> count, step -> step, it -> m, value -> t
  Returns index of leftmost matching element, or -1 if not found."""
  # Phase 1: lower_bound
  l = 0                         # first
  count = len(A)                # std::distance(first, last)
  while count > 0:
    step = count // 2           # step = count / 2
    m = l + step                # it = first; std::advance(it, step)
    if A[m] < t:
      l = m + 1                 # first = ++it
      count -= step + 1         # count -= step + 1
    else:
      count = step              # count = step
  # Phase 2: binary_search equality check
  if l < len(A) and not (t < A[l]):  # C++: !(value < *first)
    return l
  return -1
\end{lstlisting}

\subsubsection{Dafny Implementation (Formally Verified)}

The Dafny version matches the Python implementation, returning $-1$ if not found:

\begin{lstlisting}[style=dafnystyle, caption=C++ STL binary\_search in Dafny]
method bsearch_cplusplus(A: array<int>, t: int) returns (idx: int)
  requires Sorted(A)
  requires A.Length <= MAX_INT32
  ensures -1 <= idx < A.Length
  ensures idx != -1 ==> A[idx] == t
  ensures idx != -1 ==> (idx == 0 || A[idx-1] < t)  // Leftmost
  ensures idx == -1 ==> forall k :: 0 <= k < A.Length ==> A[k] != t
{
  var l := 0;
  var count := A.Length;
  while count > 0
    invariant 0 <= l <= A.Length
    invariant 0 <= count
    invariant l + count <= A.Length
    invariant forall k :: 0 <= k < l ==> A[k] < t
    invariant forall k :: l + count <= k < A.Length ==> A[k] >= t
    decreases count
  {
    var step := count / 2;
    var m := l + step;
    if A[m] < t {
      l := m + 1;
      count := count - step - 1;
    } else {
      count := step;
    }
  }
  if l < A.Length && A[l] == t {
    return l;
  }
  return -1;
}
\end{lstlisting}

\subsection{Comparison of Standard Library Styles}

\begin{center}
\begin{tabular}{lccccc}
\toprule
\textbf{Aspect} & \textbf{BSD (C)} & \textbf{glibc (C)} & \textbf{Java} & \textbf{C++ STL} \\
\midrule
Comparison & 3-way & 3-way & 3-way & 2-way \\
Bounds style & \texttt{lim} count & $[l, r)$ & $[l, r]$ & count \\
Equality & Inside loop & Inside loop & Inside loop & After loop \\
Not found & NULL & NULL & $-(l+1)$ & \texttt{false} \\
Finds & Any & Any & Any & Leftmost \\
Equivalent to & bsearch1 & bsearch1 & bsearch2 & bsearch3 \\
\bottomrule
\end{tabular}
\end{center}

\section{Combined Search: \texttt{bsearch\_ultimate}}
\label{sec:ultimate}

The preceding variants each answer a specific question: does the
target exist? Where is the leftmost/rightmost occurrence? Where would
it be inserted? We now present a single function that answers all of
these questions in one call, using two binary searches
(\texttt{bisect\_left} followed by \texttt{bisect\_right}).

\subsection{Return Convention}

The function returns a pair $(x, y)$ with three possible
interpretations:

\begin{center}
\begin{tabular}{lcl}
\toprule
\textbf{Case} & \textbf{Return} & \textbf{Meaning} \\
\midrule
Found (unique) & $(i, i)$ with $i \geq 0$ & $A[i] = t$, no duplicates \\
Found (duplicates) & $(l, r)$ with $0 \leq l < r$ & $A[l..r]$ all equal $t$ \\
Not found & $(-1, j)$ with $j \geq 0$ & $j$ is the \texttt{bisect\_left} insertion point \\
\bottomrule
\end{tabular}
\end{center}

The three cases are unambiguous: if $x = -1$ the target is absent;
otherwise $x = y$ indicates a unique match and $x < y$ indicates
duplicates. The insertion point $j$ is the smallest index such that
$A[j] \geq t$ (or $n$ if all elements are less than $t$).

\subsection{Algorithm}

We present three formulations. The \emph{self-contained} version
inlines both binary searches with a peek optimization that skips
Phase~2 for unique matches. The \emph{compositional} version delegates
to \texttt{bisect\_left} and \texttt{bisect\_right}, making the
structure transparent. The \emph{count-based} version uses the C++ STL
iteration pattern, where a \texttt{count} variable tracks the
remaining search space and strictly decreases at every step.

\subsubsection*{Version A: Self-contained with peek optimization}

\begin{algorithm}[H]
\caption{bsearch\_ultimate (self-contained)}
\begin{algorithmic}[1]
\Function{bsearch\_ultimate}{$A, t$}
    \State \Comment{Phase 1: bisect\_left}
    \State $lo \gets 0$, $hi \gets n$
    \While{$lo < hi$}
        \State $m \gets lo + \lfloor(hi - lo) / 2\rfloor$
        \If{$A[m] < t$}
            \State $lo \gets m + 1$
        \Else
            \State $hi \gets m$
        \EndIf
    \EndWhile
    \State $left \gets lo$
    \If{$left \geq n$ \textbf{or} $A[left] \neq t$}
        \State \Return $(-1, left)$  \Comment{Not found}
    \EndIf
    \If{$left = n - 1$ \textbf{or} $A[left + 1] \neq t$}
        \State \Return $(left, left)$  \Comment{Unique match --- skip Phase 2}
    \EndIf
    \State \Comment{Phase 2: bisect\_right (starting from $left + 2$)}
    \State $lo \gets left + 2$, $hi \gets n$
    \While{$lo < hi$}
        \State $m \gets lo + \lfloor(hi - lo) / 2\rfloor$
        \If{$A[m] \leq t$}
            \State $lo \gets m + 1$
        \Else
            \State $hi \gets m$
        \EndIf
    \EndWhile
    \State \Return $(left, lo - 1)$
\EndFunction
\end{algorithmic}
\end{algorithm}

\subsubsection*{Version B: Compositional via bisect\_left and bisect\_right}

\begin{algorithm}[H]
\caption{bsearch\_ultimate (compositional)}
\begin{algorithmic}[1]
\Function{bsearch\_ultimate}{$A, t$}
    \State $left \gets \Call{bisect\_left}{A, t}$
    \If{$left \geq n$ \textbf{or} $A[left] \neq t$}
        \State \Return $(-1, left)$  \Comment{Not found}
    \EndIf
    \State $right \gets \Call{bisect\_right}{A, t} - 1$
    \State \Return $(left, right)$
\EndFunction
\end{algorithmic}
\end{algorithm}

\subsubsection*{Version C: Count-based (C++ STL style)}

\begin{algorithm}[H]
\caption{bsearch\_ultimate (count-based)}
\begin{algorithmic}[1]
\Function{bsearch\_ultimate}{$A, t$}
    \State \Comment{Phase 1: lower\_bound (count-based bisect\_left)}
    \State $\mathit{first} \gets 0$, $\mathit{count} \gets n$
    \While{$\mathit{count} > 0$}
        \State $\mathit{step} \gets \lfloor\mathit{count} / 2\rfloor$
        \State $\mathit{mid} \gets \mathit{first} + \mathit{step}$
        \If{$A[\mathit{mid}] < t$}
            \State $\mathit{first} \gets \mathit{mid} + 1$
            \State $\mathit{count} \gets \mathit{count} - \mathit{step} - 1$
        \Else
            \State $\mathit{count} \gets \mathit{step}$
        \EndIf
    \EndWhile
    \State $left \gets \mathit{first}$
    \If{$left \geq n$ \textbf{or} $A[left] \neq t$}
        \State \Return $(-1, left)$  \Comment{Not found}
    \EndIf
    \If{$left = n - 1$ \textbf{or} $A[left + 1] \neq t$}
        \State \Return $(left, left)$  \Comment{Unique match --- skip Phase 2}
    \EndIf
    \State \Comment{Phase 2: upper\_bound (count-based bisect\_right, from $left + 2$)}
    \State $\mathit{first} \gets left + 2$, $\mathit{count} \gets n - (left + 2)$
    \While{$\mathit{count} > 0$}
        \State $\mathit{step} \gets \lfloor\mathit{count} / 2\rfloor$
        \State $\mathit{mid} \gets \mathit{first} + \mathit{step}$
        \If{$A[\mathit{mid}] \leq t$}
            \State $\mathit{first} \gets \mathit{mid} + 1$
            \State $\mathit{count} \gets \mathit{count} - \mathit{step} - 1$
        \Else
            \State $\mathit{count} \gets \mathit{step}$
        \EndIf
    \EndWhile
    \State \Return $(left, \mathit{first} - 1)$
\EndFunction
\end{algorithmic}
\end{algorithm}

Version~A is more efficient: it skips Phase~2 entirely for unique
matches via an $O(1)$ peek, and when duplicates are present it starts
Phase~2 from $\mathit{left}+2$ (since both $A[\mathit{left}]$ and
$A[\mathit{left}+1]$ are already confirmed). Version~B is the
clearest, relying on the correctness of \texttt{bisect\_left} and
\texttt{bisect\_right} as verified building blocks. Version~C shares
the peek optimization with~A but uses count-based iteration: the
\texttt{count} variable strictly decreases in every branch, making
termination structurally guaranteed (see Section~\ref{sec:variant5},
Performance box). In compiled languages, this is the preferred style
used by the C++ and BSD standard libraries.

\subsection{Complexity}

Each phase is $O(\log n)$, so the worst case (duplicates) is $O(\log
n)$. In Version~A, unique matches skip Phase~2 entirely via an $O(1)$
peek at $A[\mathit{left}+1]$, reducing the cost to a single
\texttt{bisect\_left}. This is particularly beneficial for datasets
with unique keys (e.g., primary key lookups), where it halves the
number of comparisons.

\subsection{Python Implementation}

\noindent\textbf{Version A: Self-contained with peek optimization}

\begin{lstlisting}[style=pythonstyle, caption=bsearch\_ultimate: self-contained with peek optimization]
def bsearch_ultimate(A: list[int], t: int) -> tuple[int, int]:
  """Combined binary search returning full match information.
  Returns: (i, i) unique, (l, r) duplicates, (-1, j) not found."""
  n = len(A)
  # Phase 1: bisect_left
  lo, hi = 0, n
  while lo < hi:
    m = lo + (hi - lo) // 2
    if A[m] < t:
      lo = m + 1
    else:
      hi = m
  left = lo
  if left >= n or A[left] != t:
    return (-1, left)               # Not found
  if left == n - 1 or A[left + 1] != t:
    return (left, left)             # Unique -- skip Phase 2
  # Phase 2: bisect_right (starting from left + 2)
  lo, hi = left + 2, n
  while lo < hi:
    m = lo + (hi - lo) // 2
    if A[m] <= t:
      lo = m + 1
    else:
      hi = m
  return (left, lo - 1)
\end{lstlisting}

\noindent\textbf{Version B: Compositional via bisect\_left and bisect\_right}

\begin{lstlisting}[style=pythonstyle, caption=bsearch\_ultimate: compositional version]
def bsearch_ultimate_v2(A: list[int], t: int) -> tuple[int, int]:
  """Compositional version using bisect_left and bisect_right.
  Returns: (i, i) unique, (l, r) duplicates, (-1, j) not found."""
  left = bisect_left(A, t)
  if left >= len(A) or A[left] != t:
    return (-1, left)               # Not found
  right = bisect_right(A, t) - 1
  return (left, right)
\end{lstlisting}

\noindent\textbf{Version C: Count-based (C++ STL style)}

\begin{lstlisting}[style=pythonstyle, caption=bsearch\_ultimate: count-based version]
def bsearch_ultimate_v3(A: list[int], t: int) -> tuple[int, int]:
  """Count-based version using C++ STL iteration pattern.
  Returns: (i, i) unique, (l, r) duplicates, (-1, j) not found."""
  n = len(A)
  # Phase 1: lower_bound (count-based bisect_left)
  first, count = 0, n
  while count > 0:
    step = count // 2
    mid = first + step
    if A[mid] < t:
      first = mid + 1
      count -= step + 1
    else:
      count = step
  left = first
  if left >= n or A[left] != t:
    return (-1, left)               # Not found
  if left == n - 1 or A[left + 1] != t:
    return (left, left)             # Unique -- skip Phase 2
  # Phase 2: upper_bound (count-based bisect_right, from left + 2)
  first, count = left + 2, n - (left + 2)
  while count > 0:
    step = count // 2
    mid = first + step
    if A[mid] <= t:
      first = mid + 1
      count -= step + 1
    else:
      count = step
  return (left, first - 1)
\end{lstlisting}

\subsection{Dafny Implementation (Formally Verified)}

All three versions share the same postconditions, ensuring they are
semantically interchangeable. Each has been verified by Dafny to
satisfy these contracts.

\medskip\noindent\textbf{Version A: Self-contained with peek optimization}

\begin{lstlisting}[style=dafnystyle, caption=bsearch\_ultimate: self-contained (Dafny)]
method bsearch_ultimate(A: array<int>, t: int)
    returns (left: int, right: int)
  requires Sorted(A)
  requires A.Length <= MAX_INT32
  // Found: valid indices, leftmost and rightmost
  ensures left >= 0 ==> 0 <= left <= right < A.Length
  ensures left >= 0 ==> A[left] == t && A[right] == t
  ensures left >= 0 ==> (left == 0 || A[left-1] < t)
  ensures left >= 0 ==> (right == A.Length-1 || A[right+1] > t)
  ensures left >= 0 ==> forall k :: left <= k <= right ==> A[k] == t
  // Not found: left == -1, right is bisect_left insertion point
  ensures left == -1 ==> 0 <= right <= A.Length
  ensures left == -1 ==> forall k :: 0 <= k < right ==> A[k] < t
  ensures left == -1 ==> forall k :: right <= k < A.Length
                           ==> A[k] > t
{
  // Phase 1: bisect_left
  var lo := 0;
  var hi := A.Length;
  while lo < hi
    invariant 0 <= lo <= hi <= A.Length
    invariant forall k :: 0 <= k < lo ==> A[k] < t
    invariant forall k :: hi <= k < A.Length ==> A[k] >= t
    decreases hi - lo
  { var m := lo + (hi - lo) / 2;
    if A[m] < t { lo := m + 1; } else { hi := m; } }
  var bl := lo;
  if bl >= A.Length || A[bl] != t { return -1, bl; }
  if bl == A.Length - 1 || A[bl + 1] != t { return bl, bl; }
  // Phase 2: bisect_right (from bl + 2)
  lo := bl + 2;  hi := A.Length;
  while lo < hi
    invariant bl + 1 < lo <= hi <= A.Length
    invariant forall k :: bl <= k < lo ==> A[k] <= t
    invariant forall k :: hi <= k < A.Length ==> A[k] > t
    decreases hi - lo
  { var m := lo + (hi - lo) / 2;
    if A[m] <= t { lo := m + 1; } else { hi := m; } }
  return bl, lo - 1;
}
\end{lstlisting}

\medskip\noindent\textbf{Version B: Compositional via bisect\_left and bisect\_right}

\begin{lstlisting}[style=dafnystyle, caption=bsearch\_ultimate\_v2: compositional (Dafny)]
method bsearch_ultimate_v2(A: array<int>, t: int)
    returns (left: int, right: int)
  requires Sorted(A)
  requires A.Length <= MAX_INT32
  // Same postconditions as Version A (omitted for brevity)
  ensures left >= 0 ==> 0 <= left <= right < A.Length
  ensures left >= 0 ==> A[left] == t && A[right] == t
  ensures left >= 0 ==> (left == 0 || A[left-1] < t)
  ensures left >= 0 ==> (right == A.Length-1 || A[right+1] > t)
  ensures left >= 0 ==> forall k :: left <= k <= right ==> A[k] == t
  ensures left == -1 ==> 0 <= right <= A.Length
  ensures left == -1 ==> forall k :: 0 <= k < right ==> A[k] < t
  ensures left == -1 ==> forall k :: right <= k < A.Length
                           ==> A[k] > t
{
  var bl := bisect_left(A, t);
  if bl >= A.Length || A[bl] != t { return -1, bl; }
  var br := bisect_right(A, t);
  return bl, br - 1;
}
\end{lstlisting}

\medskip\noindent\textbf{Version C: Count-based (C++ STL style)}

\begin{lstlisting}[style=dafnystyle, caption=bsearch\_ultimate\_v3: count-based (Dafny)]
method bsearch_ultimate_v3(A: array<int>, t: int)
    returns (left: int, right: int)
  requires Sorted(A)
  requires A.Length <= MAX_INT32
  // Same postconditions as Version A (omitted for brevity)
  ensures left >= 0 ==> 0 <= left <= right < A.Length
  ensures left >= 0 ==> A[left] == t && A[right] == t
  ensures left >= 0 ==> (left == 0 || A[left-1] < t)
  ensures left >= 0 ==> (right == A.Length-1 || A[right+1] > t)
  ensures left >= 0 ==> forall k :: left <= k <= right ==> A[k] == t
  ensures left == -1 ==> 0 <= right <= A.Length
  ensures left == -1 ==> forall k :: 0 <= k < right ==> A[k] < t
  ensures left == -1 ==> forall k :: right <= k < A.Length
                           ==> A[k] > t
{
  // Phase 1: lower_bound (count-based bisect_left)
  var first := 0;
  var count := A.Length;
  while count > 0
    invariant 0 <= first <= A.Length
    invariant 0 <= count <= A.Length
    invariant first + count <= A.Length
    invariant forall k :: 0 <= k < first ==> A[k] < t
    invariant forall k :: first + count <= k < A.Length ==> A[k] >= t
    decreases count
  { var step := count / 2;
    var mid := first + step;
    if A[mid] < t {
      first := mid + 1;
      count := count - step - 1;
    } else { count := step; } }
  var bl := first;
  if bl >= A.Length || A[bl] != t { return -1, bl; }
  if bl == A.Length - 1 || A[bl + 1] != t { return bl, bl; }
  // Phase 2: upper_bound (count-based bisect_right, from bl + 2)
  first := bl + 2;
  count := A.Length - (bl + 2);
  while count > 0
    invariant bl + 2 <= first <= A.Length
    invariant 0 <= count <= A.Length
    invariant first + count <= A.Length
    invariant forall k :: bl <= k < first ==> A[k] <= t
    invariant forall k :: first + count <= k < A.Length ==> A[k] > t
    decreases count
  { var step := count / 2;
    var mid := first + step;
    if A[mid] <= t {
      first := mid + 1;
      count := count - step - 1;
    } else { count := step; } }
  return bl, first - 1;
}
\end{lstlisting}

\subsection{Universality: Recovering All Variants}
\label{sec:universality}

A single call $\texttt{(x, y)} = \texttt{bsearch\_ultimate}(A, t)$
contains sufficient information to implement every other search variant
and derived function presented in this paper, without any additional
binary searches.

\subsubsection{Recovering bisect\_left and bisect\_right}

The two insertion points are the building blocks for all derived
functions. They can be extracted from $(x, y)$ as:

\[
\texttt{bisect\_left} = \begin{cases} x & \text{if } x \geq 0 \\[2pt] y & \text{if } x = -1 \end{cases}
\qquad
\texttt{bisect\_right} = \begin{cases} y + 1 & \text{if } x \geq 0 \\[2pt] y & \text{if } x = -1 \end{cases}
\]

For convenience, let $\mathit{bl} = \texttt{bisect\_left}$ and
$\mathit{br} = \texttt{bisect\_right}$ below.

\subsubsection{Recovering All Variants and Derived Functions}

\begin{center}
\small
\begin{tabular}{p{3.1cm}p{4.8cm}p{5.5cm}}
\toprule
\textbf{Function} & \textbf{From $(x, y)$} & \textbf{Notes} \\
\midrule
\multicolumn{3}{l}{\textit{Core search variants}} \\
\midrule
bsearch1, 2, bsd, gcc (any match)
  & \texttt{x}
  & $x \geq 0$ is a valid index; $-1$ if absent \\
bsearch3 (leftmost)
  & \texttt{x}
  & $x$ is always leftmost when found \\
bsearch4, 5 (rightmost)
  & \texttt{y if x >= 0 else -1}
  & $y$ is always rightmost when found \\
bsearch\_java
  & \texttt{x if x >= 0 else -(y+1)}
  & Encodes insertion point when absent \\
bsearch\_cplusplus
  & \texttt{x}
  & C++ lower\_bound also finds leftmost \\
\midrule
\multicolumn{3}{l}{\textit{Insertion point functions}} \\
\midrule
bisect\_left
  & $\mathit{bl} = \texttt{x if x >= 0 else y}$
  & Insert before existing entries of $t$ \\
bisect\_right
  & $\mathit{br} = \texttt{y + 1 if x >= 0 else y}$
  & Insert after existing entries of $t$ \\
\midrule
\multicolumn{3}{l}{\textit{Derived query functions}} \\
\midrule
find\_rank
  & $\mathit{bl}$
  & Count of elements $< t$ \\
find\_pred\_strict
  & $\mathit{bl} - 1$ if $\mathit{bl} > 0$, else $-1$
  & Largest index where $A[i] < t$ \\
find\_ceil
  & $\mathit{bl}$ if $\mathit{bl} < n$, else $-1$
  & Smallest index where $A[i] \geq t$ \\
find\_floor
  & $\mathit{br} - 1$ if $\mathit{br} > 0$, else $-1$
  & Largest index where $A[i] \leq t$ \\
find\_succ\_strict
  & $\mathit{br}$ if $\mathit{br} < n$, else $-1$
  & Smallest index where $A[i] > t$ \\
count duplicates
  & $y - x + 1$ if $x \geq 0$, else $0$
  & Number of occurrences of $t$ \\
\bottomrule
\end{tabular}
\end{center}

Note that \texttt{find\_range}$(a, b)$ requires two separate calls:
$\texttt{bsearch\_ultimate}(A, a)$ for $\mathit{bl}_a$ and
$\texttt{bsearch\_ultimate}(A, b)$ for $\mathit{br}_b$, yielding
count $= \mathit{br}_b - \mathit{bl}_a$.

\section{Comprehensive Comparison Table}
\label{sec:comparison}

Table~\ref{tab:comparison} summarizes the key design choices across
all variants. The ``Category'' column indicates whether equality is
checked inside the loop (3-way) or deferred (2-way). The ``Mid''
column indicates floor or ceiling division for the midpoint. BSD and
C++ use count-based iteration rather than explicit $l$/$r$ bounds.

\begin{table}[h!]
\centering
\small
\begin{tabular}{@{}p{2.0cm}cccccccc@{}}
\toprule
\textbf{Variant} & \textbf{Category} & \textbf{Bounds} & \textbf{Init $r$} & \textbf{Loop} & \textbf{Mid} & \textbf{$r$ upd} & \textbf{$l$ upd} & \textbf{Goal} \\
\midrule
bsearch1 & 3-way & $[l,r)$ & $n$ & $<$ & floor & $m$ & $m+1$ & any \\
bsearch2 & 3-way & $[l,r]$ & $n-1$ & $\leq$ & floor & $m-1$ & $m+1$ & any \\
bsearch3 & 2-way & $[l,r)$ & $n$ & $<$ & floor & $m$ & $m+1$ & leftmost \\
bsearch4 & 2-way & $[l,r)$ & $n$ & $<$ & floor & $m$ & $m+1$ & rightmost \\
bsearch5 & 2-way & $[l,r]$ & $n-1$ & $<$ & ceil & $m-1$ & $m$ & rightmost \\
bisect\_left & 2-way & $[l,r)$ & $n$ & $<$ & floor & $m$ & $m+1$ & insert pt \\
bisect\_right & 2-way & $[l,r)$ & $n$ & $<$ & floor & $m$ & $m+1$ & insert pt \\
\midrule
\multicolumn{9}{l}{\textit{Standard Library Implementations}} \\
\midrule
bsearch\_gcc & 3-way & $[l,r)$ & $n$ & $<$ & floor & $m$ & $m+1$ & any \\
bsearch\_bsd & 3-way & lim & $n$ & $\neq 0$ & floor & --- & $m+1$ & any \\
bsearch\_java & 3-way & $[l,r]$ & $n-1$ & $\leq$ & floor & $m-1$ & $m+1$ & any$^\dagger$ \\
bsearch\_cpp & 2-way & count & $n$ & $> 0$ & floor & --- & $m+1$ & leftmost \\
bsearch\_ultimate & 2-way & $[l,r)$ & $n$ & $<$ & floor & $m$ & $m+1$ & all$^\ddagger$ \\
\bottomrule
\end{tabular}
\caption{Summary of binary search variants. $^\dagger$Java also returns the bisect\_left insertion point when not found. $^\ddagger$Returns leftmost, rightmost, and insertion point via two binary searches.}
\label{tab:comparison}
\end{table}

\section{Common Pitfalls and How to Avoid Them}
\label{sec:pitfalls}

\citet{bentley1986} found that most programmers who incorrectly implemented binary search made errors in defining the exit conditions. The following are the most common bugs:

\subsection{Pitfall 1: Integer Overflow in Midpoint}
\begin{itemize}[noitemsep]
  \item \textbf{Wrong}: $mid = (l + r) / 2$ --- can overflow when $l + r > \text{INT\_MAX}$
  \item \textbf{Right}: $mid = l + (r - l) / 2$ --- always safe
\end{itemize}

\subsection{Pitfall 2: Mixing Conventions}

\textbf{Disaster recipe}:
\begin{itemize}[noitemsep]
    \item Initialize $r = n$ (exclusive convention)
    \item Use $l \leq r$ (inclusive convention)
    \item Access $A[r]$ which is out of bounds!
\end{itemize}

\textbf{Rule}: Pick one convention and stick with it throughout.

\subsection{Pitfall 3: Infinite Loop with \texorpdfstring{$l = m$}{l = m}}

When using $l \gets m$ (not $m + 1$), the floor midpoint can cause $m = l$ when $r = l + 1$, leading to no progress.

\textbf{Solutions}:
\begin{itemize}[noitemsep]
    \item Use ceiling: $m = l + \lceil(r - l) / 2\rceil$
    \item Or equivalently, with integer division: $m = l + \lfloor(r - l + 1) / 2\rfloor$
\end{itemize}

\subsection{Pitfall 4: Off-by-One in Post-Loop Check}

For leftmost/rightmost variants:
\begin{itemize}[noitemsep]
    \item Leftmost: Check $A[l]$, but first verify $l < n$
    \item Rightmost: Check $A[r-1]$, but first verify $r > 0$
\end{itemize}

\subsection{Pitfall 5: Empty Array}

Always handle $n = 0$ as a special case, especially for variants that access $A[l]$ or $A[r-1]$ after the loop.

\section{Decision Guide: Which Variant to Use?}
\label{sec:decision}

Choose a variant based on what your application requires from the
search result.

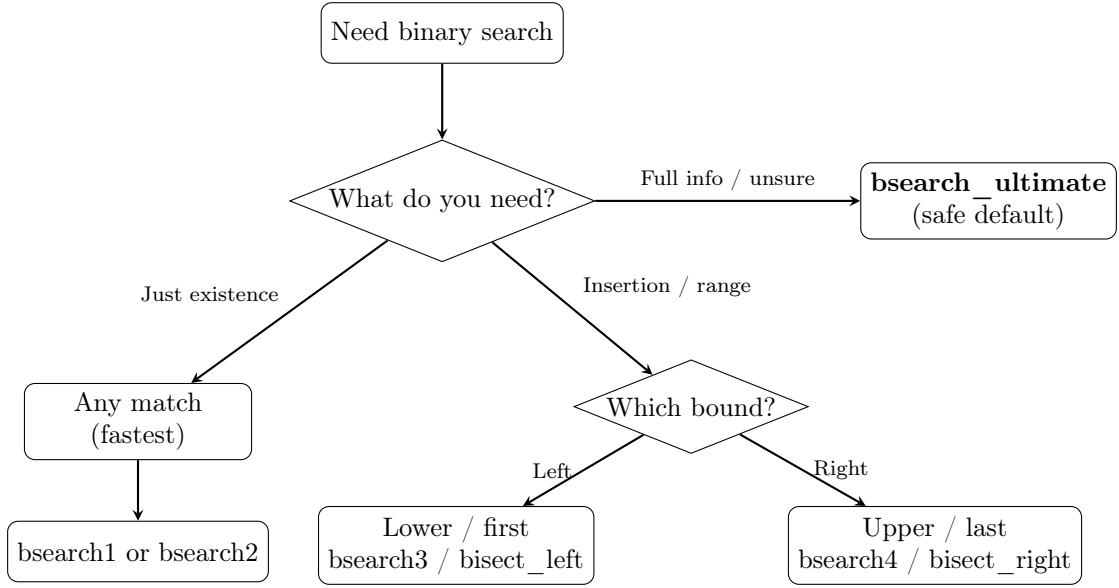
\begin{center}
\begin{tikzpicture}[
    node distance=1.8cm and 2.5cm,
    decision/.style={diamond, draw, aspect=2.5, inner sep=1pt, font=\small, minimum width=2cm},
    block/.style={rectangle, draw, rounded corners, minimum width=3cm, minimum height=0.8cm, font=\small, align=center},
    arrow/.style={->, >=stealth, thick}
]
\node[block] (start) {Need binary search};
\node[decision, below=1.0cm of start] (intent) {What do you need?};

\node[block, right=3.5cm of intent] (ultimate) {\textbf{bsearch\_ultimate}\\(safe default)};

\node[block, below left=2.0cm and 1.5cm of intent] (any) {Any match\\(fastest)};
\node[block, below=0.8cm of any] (vars12) {bsearch1 or bsearch2};

\node[decision, below right=2.0cm and 1.5cm of intent] (bound) {Which bound?};
\node[block, below left=1.0cm and 0.5cm of bound] (left) {Lower / first\\bsearch3 / bisect\_left};
\node[block, below right=1.0cm and 0.5cm of bound] (right) {Upper / last\\bsearch4 / bisect\_right};

\draw[arrow] (start) -- (intent);
\draw[arrow] (intent) -- node[above, font=\scriptsize, sloped] {Full info / unsure} (ultimate);
\draw[arrow] (intent) -- node[left, font=\scriptsize, anchor=south east] {Just existence} (any);
\draw[arrow] (intent) -- node[right, font=\scriptsize, anchor=south west] {Insertion / range} (bound);
\draw[arrow] (any) -- (vars12);
\draw[arrow] (bound) -- node[left, font=\scriptsize] {Left} (left);
\draw[arrow] (bound) -- node[right, font=\scriptsize] {Right} (right);
\end{tikzpicture}
\end{center}

\begin{enumerate}
    \item \textbf{Safe default (\texttt{bsearch\_ultimate})}: Returns
      leftmost index, rightmost index, and insertion point in a single
      call. Use when you need full match information, when
      requirements may evolve, or when duplicates may be present. It
      automatically optimizes for unique keys by skipping Phase~2 via
      an $O(1)$ peek.

    \item \textbf{Just checking existence (bsearch1, bsearch2)}: Use
      when you only need to know whether $t$ is present and any
      matching index suffices. These return immediately upon finding a
      match, making them the fastest variants. Prefer bsearch1
      ($[l,r)$ bounds) for simpler off-by-one reasoning, or bsearch2
      ($[l,r]$ bounds) if inclusive bounds feel more natural.

    \item \textbf{Specific boundaries}: Use when you need a particular
      occurrence or insertion point:
      \begin{itemize}[noitemsep]
          \item \emph{Leftmost match}: bsearch3 or bisect\_left
          \item \emph{Rightmost match}: bsearch4 or bisect\_right
          \item \emph{Insertion point}: bisect\_left (before existing
            entries) or bisect\_right (after existing entries)
      \end{itemize}

    \item \textbf{Derived queries}: For predecessor, successor, floor,
      ceiling, rank, or range counting, use the functions in
      Section~\ref{sec:derived}---thin wrappers around bisect\_left
      and bisect\_right.

    \item \textbf{Performance-critical inner loops}: When every cycle
      counts, consider Bottenbruch (bsearch5) or a count-based
      formulation (C++ STL style). These execute only one comparison
      per iteration, improving branch prediction on modern CPUs.
\end{enumerate}

\section{Testing Strategy and Validation}
\label{sec:testing}

Before formal verification, comprehensive testing provides confidence that implementations behave correctly. Our test suite (1,844 lines of Python) covers multiple dimensions of correctness.

\subsection{Test Categories}

Table~\ref{tab:test-categories} summarizes the ten test suites and
their coverage goals. Each suite targets a different dimension of
correctness, from basic functionality to cross-variant consistency
under random inputs.

\begin{table}[h!]
\centering
\small
\begin{tabular}{p{3.5cm}p{5cm}p{4cm}}
\toprule
\textbf{Test Suite} & \textbf{Purpose} & \textbf{Coverage} \\
\midrule
Basic Search & Core functionality for 8 variants (BSD/glibc/C++) & Found/not found, all positions \\
Duplicates & Leftmost/rightmost/any correctness & Arrays with multiple duplicates \\
Edge Cases & Boundary conditions & All-same-elements arrays \\
Derived Functions & All 6 query functions & rank, pred, succ, floor, ceil, range \\
Bisect Functions & Insertion point correctness & With/without target present \\
Leftmost/Rightmost Consistency & Cross-variant consistency & 50 random arrays \\
Random Stress & Correctness under random inputs & 100 arrays $\times$ 8 variants \\
Java binarySearch & Insertion point return semantics & Found, not-found, bisect\_left equivalence \\
Ultimate bsearch & All 3 versions (A/B/C): combined left/right/insertion & Unique, duplicates, edge cases, cross-validation across versions \\
Faulty Detection & Bug injection validation & 9 distinct bug types \\
\bottomrule
\end{tabular}
\caption{Test suite categories and their coverage goals.}
\label{tab:test-categories}
\end{table}

\subsection{Edge Cases Tested}

The most error-prone scenarios receive focused testing:

\begin{enumerate}[noitemsep]
    \item \textbf{Empty array} ($n = 0$): Must return $-1$ without accessing any elements
    \item \textbf{Single element}: Target equals, less than, or greater than the element
    \item \textbf{Target at boundaries}: First element, last element
    \item \textbf{Target outside range}: Smaller than all, larger than all
    \item \textbf{Target in gap}: Between consecutive elements
    \item \textbf{All same elements}: Array of identical values
    \item \textbf{Two elements}: Minimal case for interval narrowing
    \item \textbf{Power-of-two sizes}: $n = 2^k$ and $n = 2^k - 1$ (historically problematic)
\end{enumerate}

\subsection{Consistency Invariants}

For arrays with duplicates, we verify structural relationships:

\begin{align*}
\text{leftmost}(t) &\leq \text{any\_match}(t) \leq \text{rightmost}(t) \\
\text{bisect\_left}(t) &= \text{leftmost}(t) \quad \text{(when $t$ exists)} \\
\text{bisect\_right}(t) &= \text{rightmost}(t) + 1 \quad \text{(when $t$ exists)}
\end{align*}

\subsection{Bug Detection Methodology}

Each faulty implementation has a specific test designed to trigger its bug:

\begin{lstlisting}[style=pythonstyle, caption=Bug detection test pattern]
MAX_ITERATIONS = 1000  # Detect infinite loops

def test_faulty_function(A, t):
    iterations = 0
    while condition:
        iterations += 1
        if iterations > MAX_ITERATIONS:
            return -2  # Infinite loop detected
        # ... algorithm body ...
\end{lstlisting}

Return codes distinguish failure modes: $-1$ = not found (normal), $-2$ = infinite loop, $-3$ = out of bounds.

\subsection{Test Results Summary}

\begin{table}[h!]
\centering
\begin{tabular}{lccc}
\toprule
\textbf{Test Suite} & \textbf{Tests} & \textbf{Passed} & \textbf{Status} \\
\midrule
Basic Search (8 variants) & 72 & 72 & \checkmark \\
Duplicates (left/right/any) & 32 & 32 & \checkmark \\
Edge Cases (all-same) & 4 & 4 & \checkmark \\
Derived Functions (6 funcs) & 37 & 37 & \checkmark \\
Bisect Functions & 12 & 12 & \checkmark \\
Leftmost/Rightmost Consistency & 50 & 50 & \checkmark \\
Random Stress (100 $\times$ 8) & 800 & 800 & \checkmark \\
Java binarySearch & 95 & 95 & \checkmark \\
Ultimate bsearch & 8,455 & 8,455 & \checkmark \\
Faulty Bug Detection & 9 & 9 & \checkmark \\
\midrule
\textbf{Total} & \textbf{9,566} & \textbf{9,566} & \checkmark \\
\bottomrule
\end{tabular}
\caption{Test execution results. All tests pass for correct implementations; all bugs detected in faulty implementations.}
\end{table}

\begin{tcolorbox}[invbox={Testing Limitations}]
Testing can only show the \emph{presence} of bugs, not their absence
\citep{dijkstra1972}. Even with 9,566 passing tests (including
cross-validation of all three \texttt{bsearch\_ultimate} versions
across thousands of random inputs), subtle bugs could remain. This
motivates formal verification (Section~\ref{sec:dafny}), which
provides mathematical proof of correctness for all possible inputs.
\end{tcolorbox}

\section{Performance Comparison}
\label{sec:performance}

To complement the correctness results, we benchmark all variants in
both Python and C. The dual-language comparison is deliberate:
\emph{interpreter overhead masks algorithmic differences in Python},
so only compiled measurements reveal the true cost structure of each
design choice. All measurements use sorted arrays of $n = 1{,}000{,}000$
elements, searching for every element as a target ($10^6$ searches per
scenario). Python timings use \texttt{time.perf\_counter} on
Python~3.13.2; C timings use \texttt{clock\_gettime(CLOCK\_MONOTONIC)}
compiled with \texttt{-O2}.

\subsection{Methodology}

Four scenarios isolate different performance axes:

\begin{itemize}[noitemsep]
  \item \textbf{All found, unique keys}: Each target appears exactly
    once. 3-way variants may exit early; 2-way always complete
    $\lfloor\log_2 n\rfloor$ iterations. Ultimate V1/V3 run only
    Phase~1 (peek skips Phase~2).
  \item \textbf{All found, with duplicates}: Targets include repeated
    values. 3-way variants exit early upon hitting any copy; ultimate
    variants must execute both phases for every duplicate group.
  \item \textbf{All not-found}: No early exits possible; every variant
    completes the full $\lfloor\log_2 n\rfloor$ iterations.  Ultimate
    variants execute only Phase~1 (not-found detected before Phase~2).
  \item \textbf{Mixed (50\%/50\%)}: A realistic workload alternating
    found and not-found targets.
\end{itemize}

\subsection{Results}

Tables~\ref{tab:perf-python} and~\ref{tab:perf-c} report the
per-search cost ($\mu$s/search) for each variant under the four
scenarios.

\begin{table}[h!]
\centering
\small
\begin{tabular}{@{}lcccc@{}}
\toprule
 & \multicolumn{2}{c}{\textbf{All Found}} & \textbf{Not} & \\
\textbf{Variant} & \textbf{Unique} & \textbf{Dups} & \textbf{Found} & \textbf{Mixed} \\
\midrule
\multicolumn{5}{l}{\textit{3-Way Comparison (early exit on match)}} \\
\midrule
bsearch1 $[l,r)$      & 3.55 & 2.76 & 4.17 & 4.72 \\
bsearch2 $[l,r]$      & 3.83 & 3.26 & 4.59 & 3.97 \\
bsearch\_gcc           & 3.71 & \textbf{2.67} & 4.00 & 3.93 \\
bsearch\_bsd           & 4.83 & 3.46 & 5.26 & 4.92 \\
bsearch\_java          & 4.08 & 2.90 & 4.36 & 4.23 \\
\midrule
\multicolumn{5}{l}{\textit{2-Way Comparison (deferred equality)}} \\
\midrule
bsearch3 leftmost      & \textbf{3.44} & 3.97 & 4.35 & 3.43 \\
bsearch4 rightmost     & 3.49 & 3.77 & \textbf{3.78} & \textbf{3.39} \\
bsearch5 Bottenbruch   & 4.48 & 4.15 & 4.15 & 3.88 \\
bisect\_left            & 3.79 & 3.62 & 3.87 & \textbf{3.37} \\
bisect\_right           & 4.22 & 3.62 & 4.27 & 4.16 \\
bsearch\_cplusplus     & 3.92 & 3.93 & 3.86 & 4.14 \\
\midrule
\multicolumn{5}{l}{\textit{Combined Search (two-phase)}} \\
\midrule
ultimate V1 (self-cont.)& 3.90 & 7.31 & 3.97 & 3.66 \\
ultimate V2 (compositional) & 8.01 & 7.95 & 4.02 & 5.82 \\
ultimate V3 (count-based)& 4.07 & 6.90 & 3.98 & 3.93 \\
\bottomrule
\end{tabular}
\caption{Python per-search cost ($\mu$s) on $n = 1{,}000{,}000$
  sorted integers, Python~3.13.2.  Bold marks the fastest in
  each column.  Lower is better.}
\label{tab:perf-python}
\end{table}

\begin{table}[h!]
\centering
\small
\begin{tabular}{@{}lcccc@{}}
\toprule
 & \multicolumn{2}{c}{\textbf{All Found}} & \textbf{Not} & \\
\textbf{Variant} & \textbf{Unique} & \textbf{Dups} & \textbf{Found} & \textbf{Mixed} \\
\midrule
\multicolumn{5}{l}{\textit{3-Way Comparison (early exit on match)}} \\
\midrule
bsearch1 $[l,r)$      & 0.233 & 0.163 & 0.110 & 0.101 \\
bsearch2 $[l,r]$      & 0.219 & 0.153 & 0.101 & 0.093 \\
bsearch\_gcc           & 0.215 & 0.154 & 0.098 & 0.091 \\
bsearch\_bsd           & 0.200 & \textbf{0.109} & 0.092 & 0.090 \\
bsearch\_java          & 0.237 & 0.170 & 0.104 & 0.098 \\
\midrule
\multicolumn{5}{l}{\textit{2-Way Comparison (deferred equality)}} \\
\midrule
bsearch3 leftmost      & 0.236 & 0.235 & 0.101 & 0.103 \\
bsearch4 rightmost     & 0.224 & 0.228 & 0.095 & 0.097 \\
bsearch5 Bottenbruch   & 0.239 & 0.238 & 0.103 & 0.109 \\
bisect\_left            & 0.220 & 0.221 & 0.097 & 0.098 \\
bisect\_right           & 0.218 & 0.223 & 0.092 & 0.089 \\
bsearch\_cplusplus     & 0.202 & 0.204 & \textbf{0.080} & \textbf{0.087} \\
\midrule
\multicolumn{5}{l}{\textit{Combined Search (two-phase)}} \\
\midrule
ultimate V1 (self-cont.)& 0.228 & 0.582 & 0.148 & 0.110 \\
ultimate V2 (compositional) & 0.391 & 0.419 & 0.100 & 0.157 \\
ultimate V3 (count-based)& \textbf{0.199} & 0.308 & 0.085 & 0.088 \\
\bottomrule
\end{tabular}
\caption{C per-search cost ($\mu$s) on $n = 1{,}000{,}000$
  sorted integers, compiled with \texttt{-O2}.  Bold marks the
  fastest in each column.  Lower is better.  Note the $15$--$50{\times}$
  speedup over Python.}
\label{tab:perf-c}
\end{table}

\subsection{Analysis}

\paragraph{Python vs.\ C: quantitative gap.}
C is $15$--$50{\times}$ faster than Python across all variants and
scenarios, confirming that Python's interpreter overhead dominates the
$O(\log n)$ comparisons. More importantly, the \emph{rankings change
between languages}: design choices that are invisible under interpreter
overhead become decisive in compiled code.

\paragraph{3-way vs.\ 2-way comparison.}
In Python, 3-way and 2-way variants perform similarly on unique keys
because interpreter overhead dwarfs the extra branch. With duplicates,
3-way gains a modest advantage:
\texttt{bsearch\_gcc} at $2.67~\mu$s is $26\%$ faster than
\texttt{bisect\_left} at $3.62~\mu$s.

In C, this effect is amplified. On duplicates,
\texttt{bsearch\_bsd} at $0.109~\mu$s is $2.0{\times}$ faster than
\texttt{bisect\_left} at $0.221~\mu$s---the limit-based iteration
with early exit produces an extremely branch-predictor-friendly
inner loop. The 2-way variants remain insensitive to duplicates
in both languages: their times are nearly identical across the
``Unique'' and ``Dups'' columns, confirming that they always
execute the full $\lfloor\log_2 n\rfloor$ iterations regardless
of data distribution.

\paragraph{Ranking reversals: count-based iteration.}
The most striking cross-language result is the performance of
count-based variants (\texttt{bsearch\_bsd},
\texttt{bsearch\_cplusplus}, ultimate~V3).  In Python, these are
mid-pack or even slow: \texttt{bsearch\_bsd} is consistently the
\emph{slowest} single-variant ($4.8$--$5.3~\mu$s), penalized by
Python's overhead on its pointer-arithmetic-style loop.  In~C, the
same variants become the \emph{fastest}: \texttt{bsearch\_cplusplus}
leads on not-found ($0.080~\mu$s) and mixed ($0.087~\mu$s), while
\texttt{bsearch\_bsd} leads on duplicates ($0.109~\mu$s).  On unique
keys, ultimate~V3 (count-based) is the overall winner at
$0.199~\mu$s. This validates the C++ STL's design choice of
count-based iteration for \texttt{std::lower\_bound}: the pattern
eliminates one variable from the loop state and produces
straight-line code that modern compilers optimize aggressively.

\paragraph{Not-found targets.}
When no matches exist, every variant completes the full
$\lfloor\log_2 n\rfloor$ iterations. In C, the not-found scenario
produces the tightest spread ($1.85{\times}$ ratio) and the
fastest absolute times ($0.080$--$0.148~\mu$s), reflecting improved
branch prediction when every comparison takes the same path. The
consistent winner is \texttt{bsearch\_cplusplus} ($0.080~\mu$s),
whose count-based 2-way loop compiles to the tightest machine code.

\paragraph{Ultimate: peek optimization and version comparison.}
On unique keys, V1 and V3 match single-search speed in both
languages because the peek optimization skips Phase~2 after a
single $O(1)$ comparison. V2 (compositional) pays a
${\sim}2{\times}$ penalty by unconditionally running both
\texttt{bisect\_left} and \texttt{bisect\_right}.

In C, V3 (count-based) emerges as the clearly superior combined
search: it is the fastest variant overall on unique keys
($0.199~\mu$s) and the fastest among the three ultimate versions on
duplicates ($0.308~\mu$s) and not-found ($0.085~\mu$s). V1
(self-contained) shows unexpectedly high cost on duplicates in C
($0.582~\mu$s vs.\ V3's $0.308~\mu$s), likely due to less favorable
branch patterns in its index-based Phase~2 loop.

\paragraph{Practical guidance.}
For compiled code needing only one answer (leftmost, rightmost, or
any match), count-based 2-way search (\texttt{bsearch\_cplusplus})
or limit-based 3-way search (\texttt{bsearch\_bsd}) are the strongest
choices. For combined search returning all three results, V3
(count-based) is recommended: it matches or beats single-search
variants on unique keys and incurs the smallest two-phase penalty on
duplicates. V2 should be preferred only when code clarity outweighs
speed.

In interpreted languages, the rankings are flatter and the absolute
differences smaller; the choice between variants matters less than in
compiled code.

\begin{tcolorbox}[invbox={Language-Dependent Rankings}]
The Python and C benchmarks produce \emph{different rankings} for
the same algorithms. Design choices that are invisible under
${\sim}3~\mu$s of interpreter overhead per call---count-based vs.\
bounds-based iteration, limit-based pointer arithmetic---become
decisive at ${\sim}0.1~\mu$s in compiled code.  Performance-sensitive
applications should benchmark in the target language rather than
extrapolating from interpreter-level measurements.
\end{tcolorbox}

\section{Formal Verification with Dafny}
\label{sec:dafny}

\subsection{About Dafny}

Dafny \citep{dafny} is a verification-aware programming language developed at Microsoft Research. It combines a high-level programming language with a specification language and an automated theorem prover. Dafny can formally verify that implementations satisfy their specifications, including:

\begin{itemize}[noitemsep]
    \item \textbf{Preconditions} (\texttt{requires}): What must be true when a method is called
    \item \textbf{Postconditions} (\texttt{ensures}): What the method guarantees upon return
    \item \textbf{Loop invariants} (\texttt{invariant}): Properties maintained by each iteration
    \item \textbf{Termination} (\texttt{decreases}): Expressions that decrease to prove loops terminate
\end{itemize}

\subsection{Verification Methodology}

We formalized all binary search variants in Dafny with explicit loop invariants and postconditions matching the Python implementations. The verification was performed using:

\begin{verbatim}
    dafny verify --progress Batch binary_search_variants.dfy
\end{verbatim}

\subsection{Helper Lemmas for Set Cardinality Proofs}
\label{sec:dafny-lemmas}

Proving that \texttt{bisect\_left} returns the \emph{count} of elements less than $t$ requires connecting index-based reasoning to set cardinality. Dafny's verifier needs assistance with this, and naive set comprehensions like \texttt{|set k: int | 0 <= k < n|} trigger warnings about missing quantifier triggers.

\textbf{Solution:} We define the set $\{0, 1, \ldots, n-1\}$ recursively via \texttt{RangeSet}, allowing Dafny to reason inductively:

\begin{lstlisting}[style=dafnystyle, caption=RangeSet: Recursive set definition to avoid trigger warnings]
ghost function RangeSet(n: int): set<int>
  decreases n
{
  if n <= 0 then {} else RangeSet(n - 1) + {n - 1}
}

lemma RangeSetProperties(n: int)
  requires n >= 0
  ensures |RangeSet(n)| == n
  ensures forall k :: k in RangeSet(n) <==> 0 <= k < n
{
  if n == 0 { } else { RangeSetProperties(n - 1); }
}
\end{lstlisting}

For \texttt{find\_range}, which computes $|\{k : a \le A[k] \le b\}|$ as \texttt{bisect\_right(b) - bisect\_left(a)}, we need to prove that set difference preserves cardinality:

\begin{lstlisting}[style=dafnystyle, caption=SetDifferenceCardinality lemma for find\_range]
lemma SetDifferenceCardinality<T>(A: set<T>, B: set<T>)
  requires B <= A  // B is a subset of A
  ensures |A - B| == |A| - |B|
{
  if B == {} { }
  else {
    var x :| x in B;
    SetDifferenceCardinality(A - {x}, B - {x});
  }
}
\end{lstlisting}

These lemmas enable Dafny to verify the full specifications of \texttt{find\_rank} and \texttt{find\_range} including their cardinality postconditions.

\subsection{Verification Results}

Table~\ref{tab:verification-results} summarizes the Dafny verification
outcomes for all implementations. All 21 correct implementations
verified successfully, confirming that they satisfy their
specifications for all possible inputs. All 6 faulty implementations
failed verification with appropriate error messages, demonstrating
Dafny's ability to detect each category of bug.

\begin{table}[h!]
\centering
\begin{tabular}{llc}
\toprule
\textbf{Method} & \textbf{Description} & \textbf{Status} \\
\midrule
\multicolumn{3}{l}{\textit{Correct Implementations}} \\
\midrule
\texttt{bsearch1} & Half-open interval $[l, r)$ & \checkmark Verified \\
\texttt{bsearch2} & Closed interval $[l, r]$ (Bentley) & \checkmark Verified \\
\texttt{bsearch3} & Leftmost match & \checkmark Verified \\
\texttt{bsearch4} & Rightmost match & \checkmark Verified \\
\texttt{bsearch5} & Bottenbruch (ceiling midpoint) & \checkmark Verified \\
\texttt{find\_floor} & Floor (largest $\leq t$) & \checkmark Verified \\
\texttt{find\_rank} & Count elements $< t$ & \checkmark Verified \\
\texttt{find\_pred\_strict} & Strict predecessor ($< t$) & \checkmark Verified \\
\texttt{find\_succ\_strict} & Strict successor ($> t$) & \checkmark Verified \\
\texttt{find\_ceil} & Ceiling (smallest $\geq t$) & \checkmark Verified \\
\texttt{find\_range} & Count in $[a, b]$ & \checkmark Verified \\
\texttt{bisect\_left} & Leftmost insertion point & \checkmark Verified \\
\texttt{bisect\_right} & Rightmost insertion point & \checkmark Verified \\
\texttt{bsearch\_gcc} & glibc style (equiv.\ to bsearch1) & \checkmark Verified \\
\texttt{bsearch\_bsd} & BSD style (limit-based iteration) & \checkmark Verified \\
\texttt{bsearch\_java} & Java style (insertion point return) & \checkmark Verified \\
\texttt{bsearch\_cplusplus} & C++ STL (count-based leftmost) & \checkmark Verified \\
\texttt{bsearch\_ultimate} & Combined left/right/insertion (self-contained) & \checkmark Verified \\
\texttt{bsearch\_ultimate\_v2} & Combined (compositional) & \checkmark Verified \\
\texttt{bsearch\_ultimate\_v3} & Combined (count-based) & \checkmark Verified \\
\midrule
\multicolumn{3}{l}{\textit{Helper Lemmas}} \\
\midrule
\texttt{RangeSetProperties} & Set cardinality lemma & \checkmark Verified \\
\texttt{SetDifferenceCardinality} & Set difference lemma & \checkmark Verified \\
\midrule
\multicolumn{3}{l}{\textit{Faulty Implementations (expected to fail)}} \\
\midrule
\texttt{bsearch1\_faulty\_v1} & Wrong loop condition & \ding{55} Failed \\
\texttt{bsearch2\_faulty\_v1} & Wrong $r$ update & \ding{55} Failed \\
\texttt{bsearch3\_faulty\_v1} & Missing bounds check & \ding{55} Failed \\
\texttt{bsearch5\_faulty\_v1} & Floor instead of ceiling & \ding{55} Failed \\
\texttt{bsearch3\_faulty\_early\_return} & Early return (not leftmost) & \ding{55} Failed \\
\texttt{bsearch\_overflow\_bug} & Integer overflow in midpoint & \ding{55} Failed \\
\midrule
\multicolumn{3}{l}{\textit{Overflow Fix (should verify)}} \\
\midrule
\texttt{bsearch\_overflow\_fixed} & Safe midpoint calculation & \checkmark Verified \\
\bottomrule
\end{tabular}
\caption{Dafny verification results. All 21 correct implementations
  verified successfully; all 6 faulty implementations failed with
  appropriate error messages.}
\label{tab:verification-results}
\end{table}

\subsection{Errors Detected in Faulty Implementations}

Dafny precisely identified each injected bug. The table below shows the exact error messages, confirming that each bug category produces a distinct, identifiable error:

\begin{table}[h!]
\centering
\footnotesize
\begin{tabular}{p{3.2cm}p{2.8cm}p{6.5cm}}
\toprule
\textbf{Faulty Method} & \textbf{Injected Bug} & \textbf{Dafny Error Messages} \\
\midrule
\texttt{bsearch1\_faulty\_v1} & 
\texttt{l <= r} with exclusive $r$ & 
\textit{``index out of range''} at line 457, 464, 469; \textit{``decreases expression might not decrease''} at line 462 \\
\midrule
\texttt{bsearch2\_faulty\_v1} & 
\texttt{r := m} instead of \texttt{r := m - 1} & 
\textit{``index out of range''} at line 483; \textit{``decreases expression might not decrease''} at line 487 \\
\midrule
\texttt{bsearch3\_faulty\_v1} & 
No check \texttt{l < A.Length} before \texttt{A[l]} & 
\textit{``index out of range''} at lines 506, 523 (\texttt{A[l]}) \\
\midrule
\texttt{bsearch5\_faulty\_v1} & 
Floor division instead of ceiling & 
\textit{``index out of range''} at line 532; \textit{``decreases expression might not decrease''} at line 536; \textit{``loop invariant violation''} at line 537 \\
\midrule
\texttt{bsearch3\_faulty\_} \texttt{early\_return} & 
Returns \texttt{m} when \texttt{A[m]==t} & 
\textit{``index out of range''} at lines 555--556; \textit{``postcondition could not be proved''} at line 566: leftmost guarantee \texttt{(idx == 0 || A[idx-1] < t)} \\
\midrule
\texttt{bsearch\_overflow\_bug} & 
\texttt{(l + r) / 2} can overflow & 
\textit{``assertion might not hold''} at line 590: \texttt{l + r <= MAX\_INT32} \\
\bottomrule
\end{tabular}
\caption{Dafny error messages. Each bug category produces distinct errors: boundary bugs $\rightarrow$ index errors; infinite loops $\rightarrow$ termination failures; semantic bugs $\rightarrow$ postcondition failures; arithmetic bugs $\rightarrow$ assertion failures.}
\end{table}

\subsection{The Famous Integer Overflow Bug}

One of the most celebrated bugs in computer science history is the integer overflow in binary search's midpoint calculation. This bug existed in Java's \texttt{Arrays.binarySearch} from its introduction until it was discovered by \citet{bloch2006} in 2006---after roughly \emph{nine years} in the JDK.

\textbf{The Bug}: The expression \texttt{(l + r) / 2} overflows when \texttt{l + r > MAX\_INT}. For a 32-bit signed integer, this occurs when both \texttt{l} and \texttt{r} exceed approximately one billion.

\textbf{The Fix}: Use \texttt{l + (r - l) / 2} instead. Since \texttt{r - l} is always non-negative and bounded by \texttt{r}, this expression cannot overflow.

Dafny can formally verify this bug by asserting the intermediate calculation fits within bounds:

\begin{lstlisting}[style=dafnystyle, caption=Dafny catches the overflow bug]
const MAX_INT32: int := 2147483647

method bsearch_overflow_bug(A: array<int>, t: int) returns (idx: int)
  requires A.Length <= MAX_INT32
  // ... other specs ...
{
  var l := 0;
  var r := A.Length;
  while l < r {
    assert l + r <= MAX_INT32;  // ERROR: assertion might not hold
    var m := (l + r) / 2;       // Overflow possible!
    // ...
  }
}

method bsearch_overflow_fixed(A: array<int>, t: int) returns (idx: int)
  requires A.Length <= MAX_INT32
  // ... other specs ...
{
  var l := 0;
  var r := A.Length;
  while l < r {
    assert l + (r - l) / 2 <= MAX_INT32;  // Verified!
    var m := l + (r - l) / 2;              // Safe
    // ...
  }
}
\end{lstlisting}

\begin{tcolorbox}[invbox={Historical Note}]
This bug affected not only Java but also C implementations in many textbooks. Jon Bentley's original code in \textit{Programming Pearls} contained this bug. In languages with arbitrary-precision integers like Python, the bug doesn't manifest as incorrect results, but the vulnerable pattern persists in widely-used code.
\end{tcolorbox}

\subsection{Python's bisect Module: Still Vulnerable}

Remarkably, Python's standard library \texttt{bisect} module (as of
Python 3.13) still uses the overflow-vulnerable midpoint calculation:

\begin{lstlisting}[language=Python, basicstyle=\ttfamily\footnotesize, frame=single, backgroundcolor=\color{backcolour}]
# From cpython/Lib/bisect.py (current as of Python 3.13)
while lo < hi:
  mid = (lo + hi) // 2  # Vulnerable form!
  if a[mid] < x:
    lo = mid + 1
  else:
    hi = mid
\end{lstlisting}

In Python, this doesn't cause incorrect results because Python
integers have arbitrary precision. However, this pattern is dangerous
for two reasons:

\begin{enumerate}[noitemsep]
    \item \textbf{Porting risk}: Developers copying this code to C, Java, or Rust will inherit the bug
    \item \textbf{Pedagogy}: The Python standard library teaches the vulnerable pattern
\end{enumerate}

The same overflow vulnerability demonstrated in our
\texttt{bsearch\_overflow\_bug} verification applies to any
fixed-width integer implementation of this pattern.

\subsection{Analysis of Verification Results}

The verification results demonstrate several key points:

\begin{enumerate}
    \item \textbf{Boundary bugs manifest as index errors}: When loop
      conditions or bound updates are incorrect (faulty variants 1, 2,
      3), Dafny detects potential out-of-bounds array access.
    
    \item \textbf{Infinite loops manifest as termination failures}:
      When the midpoint calculation or update doesn't guarantee
      progress (faulty variants 2, 5), Dafny reports ``decreases
      expression might not decrease.''
    
    \item \textbf{Semantic bugs manifest as postcondition failures}:
      When the algorithm returns a result that doesn't satisfy the
      specification (faulty early return), Dafny reports
      ``postcondition could not be proved.''
\end{enumerate}

\begin{tcolorbox}[invbox={Formal Verification Value}]
Dafny's verification provides mathematical proof that:
\begin{enumerate}[noitemsep]
    \item All 21 binary search implementations (20 main methods plus the overflow fix) are \textbf{provably correct}
    \item Loop invariants are maintained through every iteration
    \item All loops terminate for all valid inputs
    \item All 6 injected bugs were \textbf{automatically detected}
    \item The integer overflow bug is caught via assertion checking
\end{enumerate}
This is stronger than testing, which can only show the absence of bugs in tested cases.
\end{tcolorbox}

\section{Conclusion}
\label{sec:conclusion}

This paper presented a unified treatment of binary search encompassing
five core variants, six derived query functions, four standard library
implementations, and a combined search that subsumes them all. Every
algorithm was provided as synchronized Python code, Dafny formal
proof, and pseudocode. Over 9,500 tests and 21 Dafny formal
verifications confirm correctness, while 6 deliberately faulty
implementations demonstrate Dafny's ability to detect common bug
categories, including the famous integer overflow bug that went
unnoticed for decades in production code.

Binary search's apparent simplicity belies significant implementation complexity. The key insights are:

\begin{enumerate}
    \item \textbf{Boundary conventions determine everything else}: Once you choose $[l, r)$ or $[l, r]$, the loop condition, initialization, and updates follow logically.
    
    \item \textbf{The loop invariant is your compass}: Write it down explicitly. Each iteration must preserve it.
    
    \item \textbf{Termination requires progress}: Ensure each iteration strictly shrinks $[l, r]$ or $[l, r)$. When using $l \gets m$, switch to ceiling midpoint.
    
    \item \textbf{Post-loop checks need bounds verification}: Before accessing $A[l]$ or $A[r-1]$, verify the index is valid.
    
    \item \textbf{Test edge cases religiously}: Empty array, single element, target smaller/larger than all elements, all elements equal to target.
\end{enumerate}

\begin{tcolorbox}[invbox={Final Memory Aid}]
\textbf{Exclusive $r$} $\Rightarrow$ Initialize $r = n$, loop while $l < r$, update $r = m$

\textbf{Inclusive $r$} $\Rightarrow$ Initialize $r = n-1$, loop while $l \leq r$, update $r = m-1$

When in doubt, use the $[l, r)$ convention---it's harder to make off-by-one errors.
\end{tcolorbox}

\bibliographystyle{plainnat}
\bibliography{dasdan}

\appendix

\section{Dafny Preconditions: \texttt{Sorted} and \texttt{MAX\_INT32}}
\label{sec:dafny-preconditions}

Every Dafny method in this paper carries the preconditions
\texttt{requires Sorted(A)} and \texttt{requires A.Length <= MAX\_INT32}.
The first captures the sortedness assumption that makes binary search
correct; the second bounds the array length so that the overflow-safe
midpoint $l + (r - l)/2$ cannot exceed the signed 32-bit range
(Section~\ref{sec:dafny}).

\begin{lstlisting}[style=dafnystyle, caption=Sorted predicate and MAX\_INT32 constant]
// 32-bit signed integer maximum for overflow checking.
const MAX_INT32: int := 2147483647

// Sortedness in non-decreasing order; precondition for every variant.
predicate Sorted(A: array<int>)
  reads A
{
  forall i, j :: 0 <= i < j < A.Length ==> A[i] <= A[j]
}
\end{lstlisting}

The \texttt{reads A} clause is required because the predicate
dereferences the array; without it, Dafny would reject the
definition. The use of \texttt{<=} (rather than \texttt{<}) admits
duplicates, which is essential for the leftmost/rightmost variants
(Sections~\ref{sec:variant3}--\ref{sec:variant4}).

\section{Dafny Auxiliary Functions for Cardinality Proofs}
\label{sec:dafny-auxiliary}
\label{app:auxiliary}

The Dafny specifications for \texttt{bisect\_left} and \texttt{bisect\_right} include cardinality postconditions that prove the return value equals the count of elements satisfying certain conditions. For example, \texttt{bisect\_left} ensures:

\begin{verbatim}
ensures idx == |set k | 0 <= k < A.Length && A[k] < t|
\end{verbatim}

Proving such cardinality properties requires auxiliary functions and lemmas to help Dafny's verifier connect index-based reasoning to set cardinality. This appendix documents these helper constructs.

\subsection{RangeSet: Recursive Set Definition}

Dafny's verifier warns about quantifiers like \texttt{|set k: int | 0 <= k < n|} because it cannot find a good ``trigger'' for instantiation. The solution is to define the set of consecutive integers $\{0, 1, \ldots, n-1\}$ recursively, which allows Dafny to reason inductively:

\begin{lstlisting}[style=dafnystyle, caption={RangeSet: Recursive definition of $\{0, 1, \ldots, n-1\}$}]
// Ghost function: only used for specification and proofs, not executable code
ghost function RangeSet(n: int): set<int>
  decreases n
{
  if n <= 0 then {} else RangeSet(n - 1) + {n - 1}
}
\end{lstlisting}

\textbf{Examples}:
\begin{itemize}[noitemsep]
    \item $\texttt{RangeSet}(0) = \{\}$
    \item $\texttt{RangeSet}(1) = \{0\}$
    \item $\texttt{RangeSet}(3) = \{0, 1, 2\}$
    \item $\texttt{RangeSet}(n) = \{0, 1, \ldots, n-1\}$
\end{itemize}

\subsection{RangeSetProperties: Cardinality and Membership Lemma}

This lemma establishes two key properties that allow us to connect bisect return values (indices) to set cardinalities:

\begin{enumerate}[noitemsep]
    \item $|\texttt{RangeSet}(n)| = n$ \hfill (the set has exactly $n$ elements)
    \item $k \in \texttt{RangeSet}(n) \Leftrightarrow 0 \leq k < n$ \hfill (membership is equivalent to bounds)
\end{enumerate}

\begin{lstlisting}[style=dafnystyle, caption=RangeSetProperties lemma: proves cardinality and membership]
lemma RangeSetProperties(n: int)
  requires n >= 0
  ensures |RangeSet(n)| == n
  ensures forall k :: k in RangeSet(n) <==> 0 <= k < n
{
  if n == 0 {
    // Base case: RangeSet(0) == {}, |{}| == 0
  } else {
    // Inductive step: RangeSet(n) == RangeSet(n-1) + {n-1}
    // By IH: |RangeSet(n-1)| == n-1
    // Therefore: |RangeSet(n)| == (n-1) + 1 == n
    RangeSetProperties(n - 1);
  }
}
\end{lstlisting}

\textbf{Usage in bisect\_left}: After the loop, we prove that the set of indices where $A[k] < t$ is exactly $\{0, 1, \ldots, l-1\}$:
\begin{lstlisting}[style=dafnystyle]
var s := set k | 0 <= k < A.Length && A[k] < t;
assert forall k :: 0 <= k < l ==> k in s;
assert forall k :: l <= k < A.Length ==> k !in s;
RangeSetProperties(l);
assert s == RangeSet(l);  // Therefore |s| == l
\end{lstlisting}

\subsection{SetDifferenceCardinality: Set Difference Lemma}

The \texttt{find\_range} function computes the count of elements in $[a, b]$ as:
\[
\texttt{bisect\_right}(b) - \texttt{bisect\_left}(a)
\]

To prove this equals $|\{k : a \leq A[k] \leq b\}|$, we need a lemma showing that set difference preserves the cardinality relationship when one set is a subset of another:

\begin{lstlisting}[style=dafnystyle, caption=SetDifferenceCardinality lemma for find\_range]
lemma SetDifferenceCardinality<T>(A: set<T>, B: set<T>)
  requires B <= A  // B is a subset of A
  ensures |A - B| == |A| - |B|
{
  if B == {} {
    // Base case: A - {} == A, and |A| - 0 == |A|
  } else {
    // Inductive step: pick an arbitrary element x from B
    var x :| x in B;
    // Apply IH to smaller sets
    SetDifferenceCardinality(A - {x}, B - {x});
    // Dafny can now conclude |A - B| == |A| - |B|
  }
}
\end{lstlisting}

\textbf{Usage in find\_range}: After computing \texttt{left := bisect\_left(a)} and \texttt{right := bisect\_right(b)}, we prove:
\begin{lstlisting}[style=dafnystyle]
var S_range := set k | 0 <= k < A.Length && a <= A[k] <= b;
var S_left  := set k | 0 <= k < A.Length && A[k] < a;
var S_right := set k | 0 <= k < A.Length && A[k] <= b;

// Step 1: Prove S_left is a subset of S_right
assert S_left <= S_right;

// Step 2: Prove S_range equals S_right minus S_left
assert S_range == S_right - S_left;

// Step 3: Apply the lemma
SetDifferenceCardinality(S_right, S_left);
// Now: |S_range| == |S_right| - |S_left| == right - left
\end{lstlisting}

\subsection{Why These Lemmas Are Necessary}

Dafny's SMT solver can verify many properties automatically, but cardinality reasoning over sets defined by comprehension is challenging for several reasons:

\begin{enumerate}
    \item \textbf{Trigger issues}: Set comprehensions like \texttt{|set k | P(k)|} lack natural triggers for the SMT solver to instantiate quantifiers.
    
    \item \textbf{Induction requirement}: Proving $|\{0, \ldots, n-1\}| = n$ requires induction, which must be guided explicitly.
    
    \item \textbf{Set arithmetic}: Properties like $|A - B| = |A| - |B|$ (when $B \subseteq A$) are not axiomatic in SMT theories.
\end{enumerate}

By providing recursive definitions (\texttt{RangeSet}) and inductive proofs (\texttt{RangeSetProperties}, \texttt{SetDifferenceCardinality}), we give Dafny the scaffolding it needs to verify the cardinality postconditions.

\subsection{Complete Auxiliary Code Listing}

For reference, here is the complete auxiliary code block as it appears in the Dafny source file:

\begin{lstlisting}[style=dafnystyle, caption=Complete auxiliary functions for cardinality proofs]
// =============================================================================
// HELPER LEMMAS FOR SET CARDINALITY PROOFS
// =============================================================================
// Dafny's verifier needs help connecting index-based reasoning to set
// cardinality. These lemmas bridge that gap.
// =============================================================================

// RangeSet: Recursive definition of {0, 1, ..., n-1}
ghost function RangeSet(n: int): set<int>
  decreases n
{
  if n <= 0 then {} else RangeSet(n - 1) + {n - 1}
}

// RangeSetProperties: Proves cardinality and membership for RangeSet
lemma RangeSetProperties(n: int)
  requires n >= 0
  ensures |RangeSet(n)| == n
  ensures forall k :: k in RangeSet(n) <==> 0 <= k < n
{
  if n == 0 {
  } else {
    RangeSetProperties(n - 1);
  }
}

// SetDifferenceCardinality: |A - B| == |A| - |B| when B is a subset of A
lemma SetDifferenceCardinality<T>(A: set<T>, B: set<T>)
  requires B <= A
  ensures |A - B| == |A| - |B|
{
  if B == {} {
  } else {
    var x :| x in B;
    SetDifferenceCardinality(A - {x}, B - {x});
  }
}
\end{lstlisting}

\end{document}